\documentclass[10pt,twoside]{article}

\usepackage{amsmath}
\usepackage{amssymb}
\usepackage{fancyhdr}
\usepackage[T1]{fontenc}
\usepackage[utf8]{inputenc}
\usepackage{newunicodechar}
\newunicodechar{′}{'}
\usepackage{mathtools}
\usepackage{wasysym}
\usepackage[svgnames]{xcolor}
\usepackage[paperheight=29.7cm,paperwidth=21.0cm,left=2.54cm,right=2.54cm,top=2.54cm,bottom=2.54cm]{geometry}

\newcommand\underrel[2]{\mathrel{\mathop{#2}\limits_{#1}}}
\usepackage[labelfont=bf]{caption}
\usepackage[parfill]{parskip}
\usepackage[etex=true,export]{adjustbox}

\usepackage[backend=bibtex,style=numeric-comp,sorting=none]{biblatex}
\renewbibmacro{in:}{}
\usepackage{hyperref}

\title{\LARGE Colossal transverse magnetoresistance due to nematic superconducting phase fluctuations in a copper oxide}
\date{}

\begin{document}
\maketitle

%\vspace{1\baselineskip}
Jonatan Wårdh\textsuperscript{1,$\ast$}, Mats Granath\textsuperscript{1}, Jie Wu\textsuperscript{2,$\ast$$\ast$}, A. T. Bollinger\textsuperscript{2}, Xi He\textsuperscript{3,4}, and Ivan Božović\textsuperscript{2,3,4}

\textsuperscript{1 }Department of Physics, University of Gothenburg, SE-41296 Gothenburg, Sweden
\newline
\textsuperscript{2 }Brookhaven National Laboratory, Upton, NY 11973, USA
\newline
\textsuperscript{3 }Department of Chemistry, Yale University, New Haven CT 06520, USA
\newline
\textsuperscript{4 }Energy Sciences Institute, Yale University, West Haven CT 06516, USA

\textbf{Electronic anisotropy (or `nematicity') has been detected in all main families of cuprate superconductors by a range of experimental techniques — electronic Raman scattering, THz dichroism, thermal conductivity, torque magnetometry, second-harmonic generation — and was directly visualized by scanning tunneling microscope (STM) spectroscopy. Using angle-resolved transverse resistance (ARTR) measurements, a very sensitive and background-free technique that can detect 0.5$\%$ anisotropy in transport, we have observed it also in La\textsubscript{2-x}Sr\textsubscript{x}CuO\textsubscript{4} (LSCO) for $\mathbf{0.02 \leq x \leq 0.25}$. Arguably the key enigma in LSCO is the rotation of the nematic director with temperature; this has not been seen before in any material. Here, we address this puzzle by measuring the angle-resolved transverse magnetoresistance (ARTMR) in LSCO. We report a discovery of colossal transverse magnetoresistance (CTMR) — an order-of-magnitude drop in the transverse resistivity in the magnetic field of $\mathbf{6}\,$T, while none is seen in the longitudinal resistivity. We show that the apparent rotation of the nematic director is caused by superconducting phase fluctuations, which are much more anisotropic than the normal-electron fluid, and their respective directors are not parallel. This qualitative conclusion is robust and follows straight from the raw experimental data. We quantify this by modelling the measured (magneto-)conductivity by a sum of two conducting channels that correspond to distinct anisotropic Drude and Cooper-pair effective mass tensors. Strikingly, the anisotropy of Cooper-pair stiffness is significantly larger than that of the normal electrons, and it grows dramatically on the underdoped side, where the fluctuations become effectively quasi-one dimensional. Our analysis is deliberately general rather than model-dependent, but we also discuss some candidate microscopic models including coupled strongly-correlated ladders where the transverse (inter-ladder) phase stiffness is low compared to the longitudinal intra-ladder stiffness, as well as the anisotropic superconducting fluctuations expected close to the transition to a pair-density wave state. The results provide important clues about the pseudogap state in the cuprate superconductors. }

Apart from a high superconducting critical temperature ($T_{\text{c}}$), copper oxides show other striking features \cite{keimer2015quantum}. Unlike in standard metals, in the normal state the rotational symmetry of electron fluid is spontaneously broken (‘electronic nematicity’) \cite{abdel2006anisotropic,hinkov2008electronic,daou2010broken,lawler2010intra,fujita2014simultaneous,lubashevsky2014optical,cyr2015two,zhang2017anomalous,zhao2017global,wu2017spontaneous}. Unlike in conventional superconductors, $T_{\text{c}}$ is set by the phase-ordering temperature rather than by the pairing-energy scale inferred from the measured single-particle response; consequently, above $T_{\text{c}}$ superconducting phase fluctuations abound \cite{emery1995importance,wang2006nernst,li2010diamagnetism,rourke2011phase,corson1999vanishing,bilbro2011temporal,kondo2015point,bovzovic2016dependence,zhou2019electron}. The fact that nematicity is strongly enhanced in underdoped cuprates implies that there may be interesting, yet to be revealed, connections with other unusual properties such as the pseudogap, antiferromagnetic fluctuations, and charge-, spin- or pair-density waves. Indeed, nematic order, or fluctuations near a nematic quantum critical point, has been suggested to be intimately related to high temperature superconductivity, in both cuprate and iron-based superconductors \cite{lederer2015enhancement,metlitski2015cooper,maier2014pairing,lederer2017superconductivity,fernandes2014drives,kuo2016ubiquitous}.

The possibility of electronic nematicity was first envisioned theoretically \cite{kivelson1998electronic,oganesyan2001quantum,zaanen2004duality,fradkin2010nematic,phillabaum2012spatial,beekman2017dual}. It was subsequently detected in several materials --- two-dimensional electron gas in high magnetic fields, the copper oxides, Fe-based superconductors, strontium ruthenates, and twisted bilayer graphene \cite{koulakov1996charge,lilly1999evidence,fernandes2010effects,chu2012divergent,borzi2007formation,kerelsky2019maximized,choi2019electronic}. Nematicity is unambiguously detected by measurements of various electronic properties — electron transport, thermal conductivity, Nernst effect, THz dichroism, magnetic torque, scanning tunneling microscopy maps, angle-resolved photoemission, electronic Raman scattering in the off-diagonal (crossed-polarization) geometry, second-harmonic generation, and angle-resolved transverse resistivity (ARTR) \cite{abdel2006anisotropic,hinkov2008electronic,daou2010broken,lawler2010intra,fujita2014simultaneous,lubashevsky2014optical,cyr2015two,zhang2017anomalous,zhao2017global,wu2017spontaneous}. Note that all these effects are observed in the normal state, so these materials are \textit{nematic metals}.

Rotational symmetry could also be broken in the superconducting state, in a \textit{nematic superconductor}. A prominent candidate is Bi\textsubscript{2}Se\textsubscript{3} doped by Cu, Sr, or Nb, where, as judged by the observed in-plane anisotropy of the critical magnetic field, a multicomponent superconducting order parameter breaks both the U(1) gauge symmetry and the point group symmetry of the lattice \cite{pan2016rotational,hecker2018vestigial}. In addition, in the normal state, such a superconductor can give rise to vestigial nematic order by breaking only the point group symmetry, without breaking the gauge symmetry. In the copper oxides, superconductivity is single-component (d$_{x^2+y^2}$) and does not break the crystal symmetry at the mean-field level. Nevertheless, in the presence of nematic order, as follows from the reduced symmetry, a sub-leading s-wave component will be induced in the superconducting state, shifting the gap nodes, while in the normal state nematicity would make superconducting fluctuations anisotropic.

Here we investigate this interplay between nematicity and superconducting fluctuations in La\textsubscript{2-x}Sr\textsubscript{x}CuO\textsubscript{4 } (LSCO) by studying the temperature and magnetic field dependence of the full conductivity tensor in the normal state. As is well established, superconducting fluctuations are essential to describe the conductivity in the transition region through the paraconductivity due to finite-lifetime Cooper-pairs \cite{aslamazov1968the}. We synthesize single-crystal LSCO films using atomic-layer-by-layer molecular beam epitaxy (ALL-MBE) on tetragonal LaSrAlO\textsubscript{4} (LSAO) substrates. For ARTR measurements, we lithographically pattern the films into a sunbeam arrangement of LSCO bars with different orientation with respect to the crystallographic axes. This allows for a complete mapping of the conductivity tensor, and with magnetic field, the magneto-conductivity. As shown in our earlier work we find a very strong signal of anisotropic superconducting fluctuations, consistent with nematic order, in the form of sharp peaks in the transverse resistivity that are systematically modulated in sign and magnitude according to the orientation of the bars \cite{wu2017spontaneous}. We now show that t his signal is rapidly suppressed by a magnetic field, the \textit{colossal transverse magnetoresistance}. The observation is consistent with the broadening of the transition region observed in the longitudinal conductance and a more gradual decrease of paraconductivity with increasing temperature, thus confirming the superconducting origin of the peaks. The principal axes of the conductivity are not aligned with a symmetry axis of the tetragonal crystal, which indicates a general in-plane nematic order with independent $x^2-y^2$(\textit{B}\textsubscript{1g}) and $xy$ (\textit{B}\textsubscript{2g}) Ising (\textit{Z}\textsubscript{2}) nematic fields. This observation of two symmetry breaking nematic orders is consistent with observations in bulk LSCO from neutron scattering of both bond-aligned and diagonal spin-density wave fluctuations \cite{wakimoto1999observation,matsuda2008magnetic}.

Even more unexpected, the anisotropy of the Cooper-pair mass (superfluid stiffness), deduced by analyzing the conductivity tensor, greatly exceeds, and is not aligned with, that of the effective mass of normal electrons. The first result follows from a simple two-fluid model of normal electron Drude conductivity and Aslamazov-Larkin paraconductivity, as a prerequisite for generating the sharp peaks observed in the transverse resistance. The second result follows directly from the observation that the principal axes of the conductivity rotate back to the high-temperature orientation when the paraconductivity is suppressed by a magnetic field.

The conclusion that the Drude and Cooper-pair mass tensors are independent challenges the Bose-Einstein condensation (BEC) interpretation of the pseudogap in the cuprate superconductors, as the anisotropy of phase stiffness would be expected to reflect the anisotropy of the effective mass of constituent electrons. The fact that the observed anisotropy of the superconducting fluctuations is extremely large, even quasi-1D in the underdoped samples, gives additional evidence of an exotic normal state. In fact, the findings are broadly reminiscent of models for high-temperature superconductivity based on coupled, strongly-correlated (Hubbard or t-J) ladders that have a low transverse inter-ladder stiffness due to pair-hopping between ladders, and a large intra-ladder pairing (pseudo) gap \cite{emery1993frustrated,arrigoni2004mechanism,fradkin2015colloquium}. Another possible interpretation is that we observe fluctuations of a pair-density wave order \cite{agterberg2020physics,berg2007dynamical,hamidian2016detection,edkins2019magnetic}, for which a difference in transverse and longitudinal stiffness would be expected. However, neither of these models seem to capture the complete physics. In the ``ladder'' scenario it may be expected that the normal transport would also be quite anisotropic and in the PDW scenario the divergence of superconducting fluctuations approaching $T_{\text{c}}$ would seem to require fine-tuning of the finite momentum and homogeneous components. In addition, the existence of both bond-aligned and diagonal nematic fields would have to be considered.

It should be emphasized that the large stiffness anisotropy that we have observed does not imply a large gap anisotropy. As is well known for the underdoped cuprates, the stiffness is low and not directly related to the superconducting gap. The stiffness quantifies slow deformations of the order parameter, and it describes the dynamic properties. The superconducting gap quantifies local pair-breaking excitations, and it is a static (mean-field) property. In fact, assuming that the pseudogap is related to pairing, this dichotomy is consistent with our observation of small normal-electron anisotropy and large pair-fluctuation anisotropy in the pseudogap regime. This crucial distinction between probing static and dynamic properties, together with the sensitive nature of the nematic order, may explain why these important observations have not been made previously.

\section{Results and Discussion}

To provide the context we begin with a summary of the general properties of the in-plane conductivity, with or without magnetic field. The conductivity takes the form of a 2D tensor which can be decomposed in accordance with its transformation under rotation:
\begin{equation}
\overline{\sigma } =
\begin{bmatrix}
\sigma_{aa} & \sigma_{ab} \\ 
\sigma_{ba} & \sigma_{bb} \\ 
\end{bmatrix}
\equiv
\begin{bmatrix}
\sigma_{0}+\sigma_{S}^{(1)} & \sigma_{\text{H}}+\sigma_{S}^{(2)} \\ 
-\sigma_{\text{H}}+\sigma_{S}^{(2)} & \sigma_{0}-\sigma_{S}^{(1)} \\ 
\end{bmatrix} 
\end{equation}
here expressed in the $a,b$ crystal frame (the [100] and [010] directions). The trace $ 2\sigma_{0}$ and the Hall conductivity $\sigma_{\text{H}}$ are left invariant under rotation, while 
$\overline{\sigma }_{S} 
=\begin{bmatrix}
\sigma_{S}^{(1)} & \sigma_{S}^{(2)} \\ 
\sigma_{S}^{(2)} & -\sigma_{S}^{(1)} \\ 
\end{bmatrix}$
, transforms as a traceless symmetric tensor. A current $\vec{J}$ running at an angle $\phi$ (with respect to the crystal axis $a$) will in general yield both a longitudinal resistivity $\rho_{\text{L}}(\phi) = \vec{E} \cdot \vec{J} / |\vec{J}|^2$ and a transverse resistivity $\rho_{\text{T}}(\phi) = \hat{z} \cdot \vec{E}\times \vec{J} / |\vec{J}|^2$, given by:
\begin{eqnarray}
%\begin{array}{c}
% 
\rho_{\text{T}}(\phi) & = &\frac{\sigma_{\text{H}}+\sqrt{\vert \text{det}\left(\overline{\sigma }_{S}\right)\vert }\sin (2(\phi -\alpha))}{\sigma_{\text{H}}^{2}+\sigma_{0}^{2}+\det\left(\overline{\sigma }_{S}\right)} \\ 
\rho_{\text{L}}(\phi) & = & \frac{\sigma_{0}+\sqrt{\vert \text{det}\left(\overline{\sigma }_{S}\right)\vert }\cos (2(\phi -\alpha))}{\sigma_{\text{H}}^{2}+\sigma_{0}^{2}+\det\left(\overline{\sigma }_{S}\right)}. 
%
%\end{array}
\end{eqnarray}
Here, $\alpha  =\frac{1}{2}\arctan\left(\sigma_{S}^{(2)}/\sigma_{S}^{(1)}\right)$, is the angle, measured from the crystal $a$-axis, along which the longitudinal resistivity is maximal. 

The traceless symmetric part of the conductivity tensor $\overline{\sigma }_{S}$ is a nematic order parameter. In fact, $\sigma_{S}^{(1)}$ ($\sigma_{S}^{(2)}$) is an order parameter for \textit{B}\textsubscript{1g} (\textit{B}\textsubscript{2g}) nematic order, given that the crystal is tetragonal. In the absence of a magnetic field ($\sigma_{\text{H}} = 0$), the appearance of a non-zero transverse voltage in zero magnetic field is thus coincident with the development of nematic order. An out-of-plane magnetic field induces an antisymmetric Hall component $\sigma_{\text{H}}$ and will give rise to a transverse resistance even without $\overline{\sigma }_{S}$. However, since $\sigma_{\text{H}}$ is rotationally invariant, the Hall resistivity $\rho_{\text{H}}(B)$ can generally be distinguished from the traceless symmetric part if the full $\sin (2(\phi -\alpha))$ angular dependence is measured and resolved. 

\begin{figure}[ht]
\centering
\includegraphics[width=0.9\textwidth]{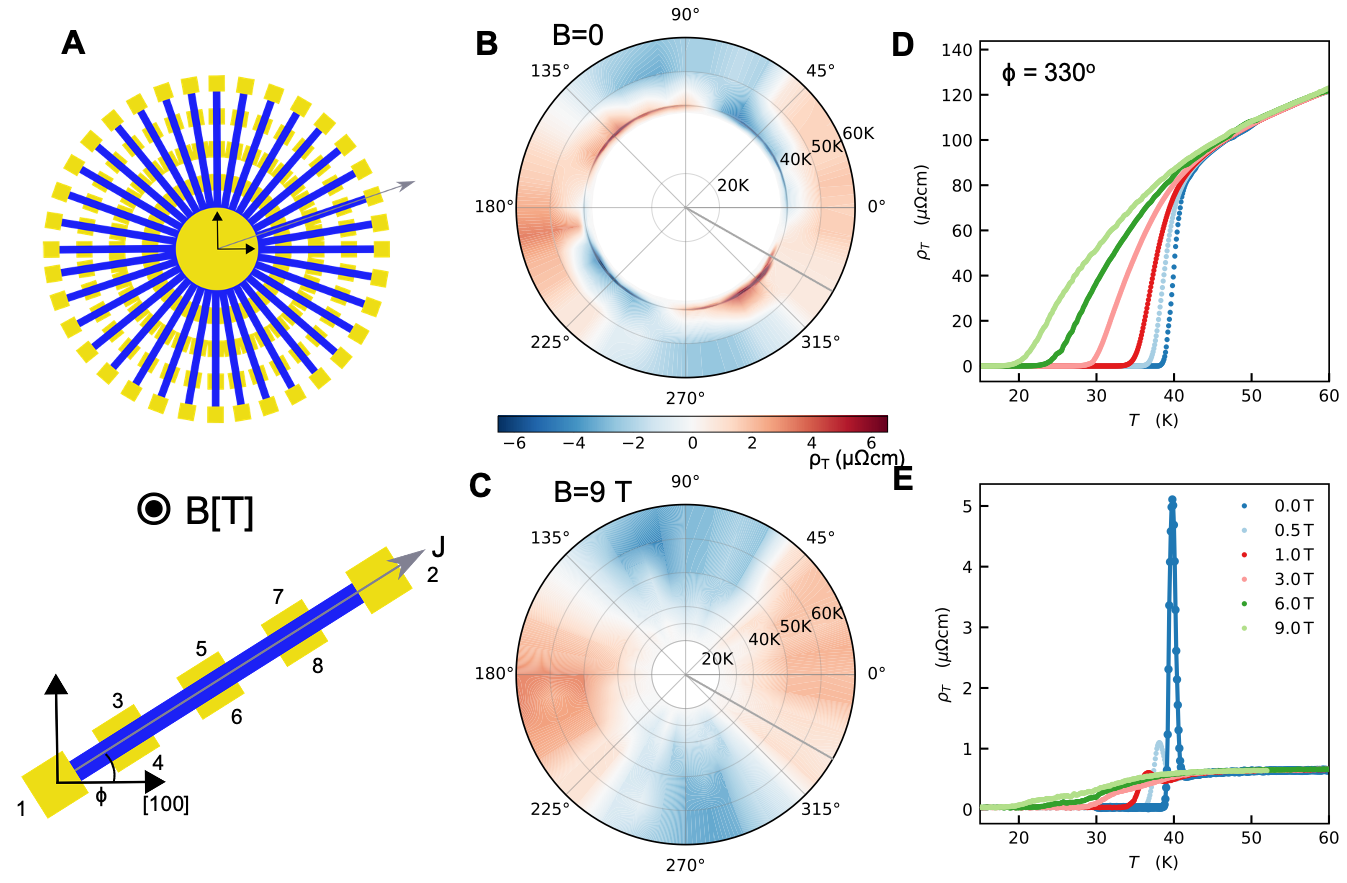}
\caption{ \label{fig_1} \textbf{ Transverse magnetoresistivity in an optimally-doped ($\mathbf{x= 0.16}$) LSCO film.} A, 36 bars of LSCO arranged in a sunflower pattern with a magnetic field perpendicular to the plane. The probe current runs along the length of the bar (blue) with gold contacts (yellow). The longitudinal voltage is measured over contacts 3-5, 4-6, 5-7 and 6-8 and averaged, while the transverse voltage is an average over 3-4, 5-6 and 7-8; $\phi$ is the angle between the current and the crystallographic [100] direction. B and C transverse resistivity in absence and presence of a magnetic field ($9\,$T) respectively, shown as a function of temperature (radial direction) and angle. The data are interpolated from the transverse resistivity measured in each bar. In D and E the longitudinal and transverse resistivities along a fixed, $\phi= 330^{\circ}$, direction is presented. In absence of a magnetic field the principal direction of the conductivity twists as a function of temperature (B) and there is a pronounced peak in the transverse resistivity (E). In the presence of a $B =9\,$T magnetic field, the transition is broadened (D), while the peak in transverse resistivity is successively suppressed (E), as the twist of the director unwinds (C), indicating that the peak is of a superconducting origin.}
\label{fig:temp}
\end{figure}

A series of experiments were performed consisting of measuring the longitudinal and transverse voltage versus current in mesoscopic Hall bar structures of LSCO manufactured from single-crystal films, synthesized by ALL-MBE on tetragonal LSAO substrates. In zero field, the data coincide with those in Ref. \cite{wu2017spontaneous}, but the experimental set-up is now upgraded to allow an out-of-plane magnetic field, up to $9\,$T, to be applied, enabling us to suppress superconductivity. The crucial aspect of the experiment is the sunbeam arrangement of Hall bars, with an orientation that varies sequentially with respect to the crystal axes, allowing for angle-resolved measurements, see Figure \ref{fig_1}A. 

Figure \ref{fig_1} provides a summary of the experimental findings near the optimal doping ($x=0.16$). (Data for additional doping levels are presented in the Supplemental Information (SI).) The Hall resistance has been identified from symmetry and subtracted from the data. The polar color plots give a representation of the magnitude and sign of the transverse resistivity as a function of temperature (radial) and orientation of the bar. The plots are based on interpolating high-resolution temperature data from some bars and discrete fixed temperature data for all bars (see Figure \ref{fig_S1}A). There is evident $\sin(2(\phi -\alpha))$ periodicity (Figure \ref{fig_1}B and Figure \ref{fig_S1}B and Figure \ref{fig_S3}) with respect to the azimuth angle $\phi$ consistent with nematic order. Most notable are the sharp peaks close to $T_{\text{c}}$ in the zero-field data, which also show the $\sin (2(\phi -\alpha))$ dependence. The orientation of principal axes of conductivity (identified as the directions along which $\rho_{\text{T}}(\phi)$ changes sign) is temperature dependent, with a rapid twist close to $T_{\text{c}}$, which coincides with the onset of the peaks. A magnetic field of $9\,$T broadens the superconducting transition and eliminates the peaks in transverse resistivity. Notably, this also unwinds the twist of the nematic director, which now shows just a single, fixed (i.e. temperature-independent) orientation, from room temperature down to $T= 0.3\,$K, the same as the high-temperature orientation in zero field. 

In Figure \ref{fig_1}D-E, data are shown along one azimuthal direction for the same sample, for a range of magnetic fields. As the field is increased, in copper oxides the superconducting transition broadens and fans out — the onset of the transition does not change much while the offset $T_{\text{c}}(R=0)$ decreases fast until a finite resistivity appears, which then grows until it saturates when the superconducting fluctuations are completely quenched. This phenomenon is well-known from the literature \cite{tinkham1988resistive} and indeed this is what we observe, as illustrated in Figure \ref{fig_1}D. The key new experimental observation here is that, concomitantly, the sharp peak in transverse resistivity is suppressed and broadened with increasing field. As can be seen in Figure \ref{fig_1}E, in the magnetic field of 6T or higher, the transverse resistivity drops by an order of magnitude. This is comparable to the colossal magnetoresistance (CMR) effect seen in the canonical CMR material La\textsubscript{0.77}Sr\textsubscript{0.33}MnO\textsubscript{3}, in the longitudinal resistivity channel \cite{schiffer1995}. The main difference and novelty here is that the `colossal' drop is seen in the transverse resistivity channel, with the maximal effect at about $T=41\,$K, while at the same temperature there is hardly any change in the longitudinal resistivity. Hence, this new effect can be called colossal transverse magnetoresistance (CTMR). 

In order to gain some deeper insights from these experimental observations we have modelled the conductivity using a rudimentary ``two-fluid'' model of normal-electron Drude conductivity and Cooper-pair normal state paraconductivity
\begin{equation}
    \overline{\sigma } =\overline{\sigma }_{\text{n}}+\overline{\sigma }_{\text{p}}
\end{equation}
The normal component is given by an anisotropic Drude conductivity
\begin{equation}
\overline{\sigma }_{\text{n}} = ne^{2}\tau \overline{m}_{\text{n}}^{-1}
\underrel{\text{princ.}}{=}   ne^{2}\tau 
\begin{bmatrix}
1/m_{\text{n},x} & 0 \\ 
0 & 1/m_{\text{n},y} \\ 
\end{bmatrix}
\end{equation}
where $n$ is the density of normal electrons, $\tau$ the (temperature-dependent) relaxation time, and $\overline{m}_{\text{n}}$ is the single-particle mass-tensor. The last equality gives the expression in the principal frame of the Drude mass tensor. To describe the contribution from fluctuating superconductivity, the so-called paraconductivity, we use a generalized Aslamazov-Larkin (AL) expression \cite{aslamazov1968the} obtained by considering an anisotropic 2D Ginzburg-Landau free energy:
\begin{equation}
\int_{}^{} a\left(T-T_{\text{c}}\right)\vert \Delta \vert^{2}+\left(\frac{\hbar^{2}}{2\overline{m}_{\text{p}}}\right)_{ij}\left(D_{i}\Delta\right)^{\ast }\left(D_{j}\Delta\right).
\label{eq:GL}
\end{equation}
Here, $ D_{i} = \partial_{i}-(2e/\hbar)A_{i}$ is the gauge invariant derivative which includes the vector potential $\vec{A}$, $\hbar$ is the Planck constant, $\overline{m}_{\text{p}}$ is the Cooper-pair mass tensor, and $a$ is a constant with the dimension energy/temperature. The last term describes the electro-magnetic response and represents the stiffness to deformation. Given \eqref{eq:GL}, we find the (generalized) AL contribution to paraconductivity (see SI)
\begin{equation}
\overline{\sigma }_{\text{p}} =\overline{\sigma }_{\text{p},0}\frac{T_{\text{c}}}{T-T_{\text{c}}}\sqrt{\det\left(\overline{m}_{\text{p}}\right)}\overline{m}_{\text{p}}^{-1}
\underrel{\text{princ.}}{=}
\sigma_{\text{p},0}\frac{T_{\text{c}}}{T-T_{\text{c}}}
\begin{bmatrix}
\sqrt{m_{\text{p},y}/m_{\text{p},x}} & 0 \\ 
0 &\sqrt{m_{\text{p},x}/m_{\text{p},y}} \\ 
\end{bmatrix}
\label{eq_6}
\end{equation}
where $\overline{\sigma }_{\text{p},0} =\frac{e^{2}}{16\hbar d}$, and $d$ is the CuO$_2$ interlayer distance. The last expression holds in the principal frame of the Cooper-pair mass tensor. The basic physics behind this paraconductivity is that the electric field induces a non-equilibrium distribution of Cooper-pairs, with a relaxation time that diverges as $ T_{\text{c}}$ is approached. The Drude part will dominate the conductivity except very close to $ T_{\text{c}}$ where the diverging paraconducting contribution gives the characteristic rapid drop in the longitudinal resistivity. This in-plane paraconductivity of cuprate superconductors has been studied earlier \cite{carballeira2001paraconductivity,curras2003plane}, but not in terms of the transverse conductance which is our main focus here.

In tetragonal cuprates, there are two distinct symmetry-allowed nematic fields: a bond-nematic, (\textit{B}\textsubscript{1g}) field transforming as $x^2-y^2$, and a diagonal-nematic, (\textit{B}\textsubscript{2g}) field transforming as $xy$. These fields break the tetragonal symmetry, and if only one of them is present it introduces a preferred direction, either along a crystal axis, or along a diagonal. If both fields are present, they will couple independently to different tensorial physical quantities such that their principal axes will not be aligned with each other or with a high-symmetry direction. The fact that the experimentally observed conductivity is in general not aligned with a high-symmetry tetragonal direction implies that both nematic fields are present. The relevant tensorial objects in the two fluid model are the Drude and Cooper-pair mass tensors, that are quantified by their mass anisotropies and principal-axes directors. Specifically, $\overline{m}_{\text{n}}$ and $\overline{m}_{\text{p}}$ are assumed to be aligned with the measured $\rho_{\text{T}}$ at the high-temperature tail, and just above $ T_{\text{c}}$, respectively. In their respective principal frames of reference, the mass anisotropies are denoted $ m_{\text{n},y}/m_{\text{n},x}$ and $ m_{\text{p},y}/m_{\text{p},x}$. We will assume these tensors to be fixed, in magnitude and orientation, such that the only temperature dependence is built into the Drude relaxation time (to fit the high temperature tail) and the intrinsic $ 1/(T-T_{\text{c}})$ temperature dependence of the AL expression. In addition, we will assume that pre-factor $\overline{\sigma }_{\text{p},0}$ may deviate from the universal expression, possible reflecting non-BCS Cooper-pair relaxation. The details of the modelling are presented in the SI. As described there, we have made several independent fits of the key quantities, all pointing to the same conclusion of effective decoupling of superconducting fluctuations from the normal electrons with a very large Cooper-pair mass anisotropy. 

\begin{figure}[ht]
\centering
\includegraphics[width=0.9\textwidth]{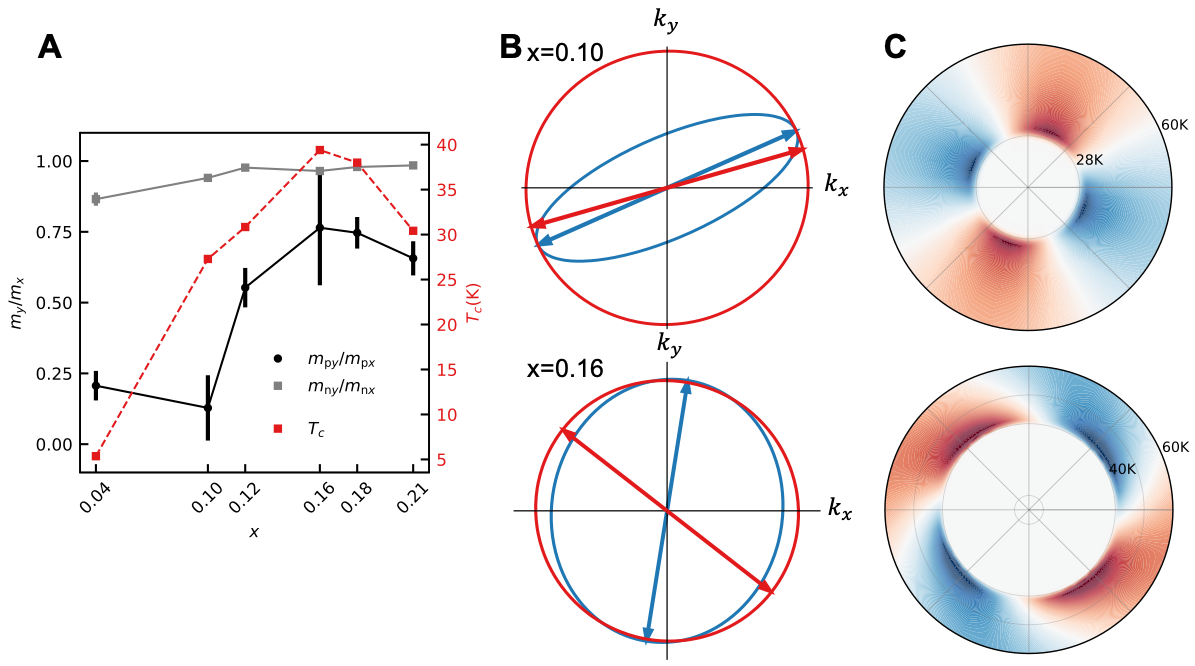}
\caption{\label{fig_2} \textbf{Modelling the zero-field conductivity.} A, Drude effective mass anisotropy and Cooper-pair mass anisotropy versus doping. B, Mass anisotropies visualized as constant energy cuts of parabolic dispersions for Cooper-pairs (blue) and normal electrons (red). C, Transverse resistivity (scale and orientation as in Figure \ref{fig_1}B) as a function of temperature and orientation with respect to crystal axes, showing the rapid rotation close to $T_{\text{c}}$ as an effect of the paraconducting contribution. The white transition between blue and red correspond to $\rho_{\text{T}} = 0$, i.e., the principal axes of conductivity. Compared to experimental data in Figure \ref{fig_1}D and E.}
\label{fig:temp2}
\end{figure}

\begin{figure}[ht]
\centering
\includegraphics[width=0.9\textwidth]{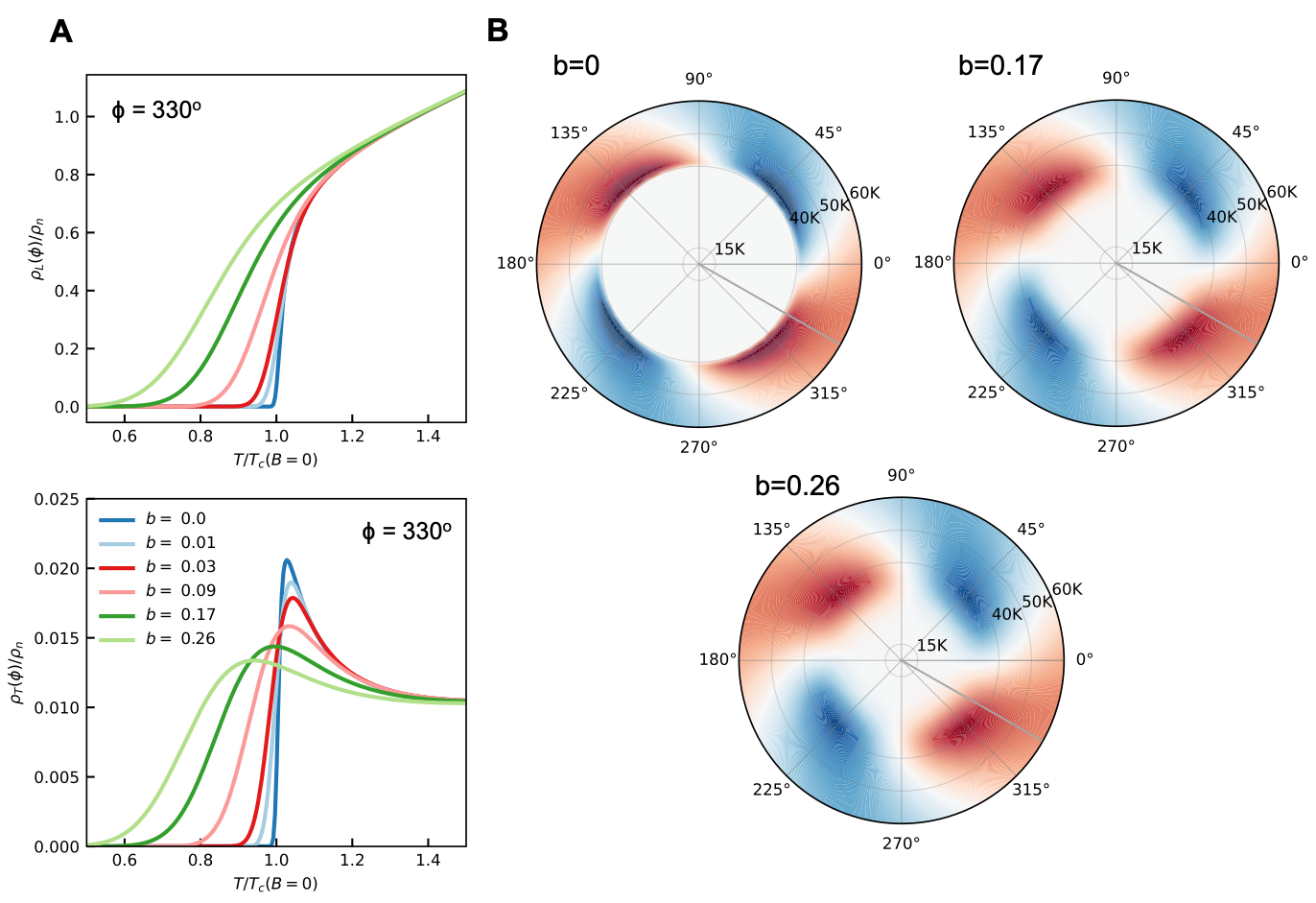}
\caption{\label{fig_3} \textbf{Modelling the magnetic field dependence of longitudinal and transverse resistivity of LSCO with $\mathbf{x= 0.16}$.} A, The calculated longitudinal and transverse resistivity $\rho_{\text{L}}(T)$ and $\rho_{\textbf{T}}(T)$, for different values of $B$ up to $9\,$T. The color code is the same as in Figure \ref{fig_2} and indicates the value of magnetic field. A The simulated data for a fixed direction, $\phi=330^{\circ}$ as the principal axis of conductivity rotates by $\phi=60^{\circ}$ from $T_{\text{c}}$ to $296\,$K. (See Figure \ref{fig_S3} and Table \ref{tab_S1}.) Here the same conditions are used with $\rho_{\text{n}}=\rho_{\text{L}}(1.3T_{\text{c}})= 107\,$cm and $ T_{\text{c}}(\text{B}=0) = 40\,$K. B, The simulated transverse resistivity in a polar-plot representation.}
\label{fig:temp3}
\end{figure}

Figure \ref{fig_2} exemplifies the main results of the modelling of the zero-field conductivity. The deduced Cooper-pair mass anisotropy is found to roughly follow the superconducting transition temperature. At low doping, $x = 0.10$, the anisotropy is estimated to be at least a factor ten, indicating that superconducting fluctuations are effectively quasi-one-dimensional. Near optimal doping, $x=0.16$, Drude and Cooper-pair mass tensors are not aligned, which causes a rapid rotation of the principal axis of conductivity given by $\rho_{\text{T}} = 0$ (white in panel C). Figure \ref{fig_3} exemplifies how the model, when extended to include the effects of the magnetic field on the paraconductivity and a phenomenological description of the broadened transition region, qualitatively reproduces the characteristic experimental observations, including the suppression of the peak in transverse resistivity and the unwinding of the sharp twist of the principal axes of conductivity.

\section{Conclusions and Outlook}

We have shown that detailed measurements of longitudinal and transverse resistivity in the normal state of LSCO, that indicate electronic nematicity, can be explained within a two-fluid model of normal electrons and finite-lifetime Cooper-pairs. A sharp peak in the transverse resistivity close to $T_{\text{c}}$ is a signature of the relaxation time of the Cooper-pairs that diverges as $T\rightarrow T_{\text{c}}$. This explanation of the peak origin is supported by our measurements which show that the peak in transverse resistivity is rapidly suppressed by an out-of-plane magnetic field that has little effect on the longitudinal resistivity at the onset temperature, an effect that we have dubbed colossal transverse magnetoresistance. The concomitant reorientation of the nematic director is also consistent with the suppression of the contribution from superconducting fluctuations. The systematic angular dependence of the peak is consistent with an underlying nematic order, and it rules out the possibility that the peak could originate from the sample inhomogeneity\footnote{When measuring resistivity in a granular system one may in general expect peaks due to grains not becoming superconducting at the same temperature. This is perhaps most easily understood by considering a network of resistors. Whenever a voltage drop is measured between two points in the grid that does not coincides with the source and drain of current the effective resistance will in general depend on the difference of resistivity between resistors in the grid. This leads to an enhancement of resistivity when one resistor goes superconducting before another \cite{nordstroem1992resistance}. However, such a peak in the transverse resistance would have an arbitrary sign consistent with the stochastic nature of the mechanism, in contrast to what is observed in single-crystal LSCO films grown by ALL-MBE \cite{wu2017spontaneous}.}.

The most striking result of the study is that the Cooper-pair mass anisotropy (or phase stiffness) is very large (more than 10, in underdoped LSCO with $x=0.10$), and apparently decoupled from a much smaller anisotropy of the mass of normal electrons. We emphasize (for details, see SI) that this unexpected behavior follows from even the simplest model of normal conductivity and Aslamazov-Larkin pair conductivity; for a peak in transverse resistivity to appear it is necessary that the pair anisotropy is substantially larger than the normal anisotropy. This is further supported by extracting the mass anisotropies from the asymptotic $T\rightarrow T_{\text{c}}$ behavior, as well as from more elaborate fits to the $T$-dependence of the data in the entire experimentally-accessible temperature range, which all independently turn out similar estimates of the mass anisotropies. 
The nature of the underlying nematicity is unknown, but we stress that it is manifested primarily via the collective response of the superconductor. The later points at some highly anisotropic superconducting state without global phase coherence, a 'phase-stiffness nematic', which is possibly even quasi-one-dimensional. Such a dichotomy between single-particle and pair response is expected neither in the standard BCS nor in common Bose-Einstein condensation scenarios, for which the phase stiffness anisotropy should simply reflect the anisotropy of effective mass of constituent electrons. This is an important new insight into the nature of cuprate superconductivity. Our data, and notably the facts that the crystal structure of the films (and the substrates they are epitaxially anchored to) is tetragonal, that the director is not aligned with the crystal axes and even rotates with temperature, and that the anisotropy of normal electrons persists up to room temperature, all point to electronic nematicity. However, that is not crucial for the above argument. Regardless of the origin of the anisotropy of the normal electron fluid, the much greater anisotropy of the superfluid stiffness is anomalous and unexpected within the standard theory of superconductivity. 

In theory, such a decoupling of single particle and collective pair-response has already been contemplated. A rare tractable model is a coupled array of strongly-correlated ladders \cite{kivelson1998electronic,emery1993frustrated,arrigoni2004mechanism,fradkin2015colloquium,zachar1998landau,granath2001nodal,carlson2002concepts}
%\textsuperscript{27, 42-44, 58-60}
. In the simplest scenario, each ladder is a Luther-Emery liquid \cite{luther1974backward} with a spin gap corresponding to singlet pairing, but no charge gap, and a diverging susceptibility (with temperature) to both superconducting and charge-density-wave orders. Single-particle tunneling between ladders can be neglected, but Josephson pair-hopping may stabilize a 2D superconductor at the temperature scale of phase-ordering between ladders. In this model, the pairing (pseudo) gap, due to breaking of local singlet pairs, can indeed be isotropic (except for the form factor, which could be d-wave). In contrast, the phase stiffness can be very anisotropic, since the intrinsic stiffness of deformations within a strongly correlated ladder can indeed be quite different from the stiffness due to Josephson coupling between ladders. This anisotropic phase stiffness would manifest itself in the pseudogap phase of the model, as highly anisotropic superconducting fluctuations. Related models have also found that the nodal ``Fermi arc'' electrons can be largely unaffected by quasi-1D charge- and spin-orders as well as superconducting fluctuations \cite{salkola1996implications,granath2002distribution,carlson2000dimensional,granath2006nodal,granath2010modeling}.
%\textsuperscript{62-66} 
Although the model of coupled ladders cannot capture the complex physics of the cuprate superconductors, unidirectional spin- and charge-density-wave orders as well as the nematic order \cite{keimer2015from,lawler2010intra,fujita2014simultaneous,ghiringhelli2012long,zheng2017the}, corresponding to broken translational and/or rotational invariance \cite{kivelson1998electronic,kivelson2003how,vojta2009lattice}, appear to be a ubiquitous phenomenon in cuprates, and may be linked to the pseudogap \cite{lederer2017superconductivity,kivelson2019linking,mukhopadhyay2019evidence}.

Recently, pair density wave (PDW) order of spatially modulated superconductivity has also emerged as one of the key new concepts in the field of high-$T_{\text{c}}$ and strongly-correlated superconductivity \cite{agterberg2020the,berg2007dynamical,hamidian2016detection,edkins2019magnetic,himeda2002stripe,berg2009theory,zelli2011mixed,baruch2008spectral,lee2014amperean,waardh2018suppression,agterberg2008dislocations,babaev2004phase}.
%\textsuperscript{45-48,73-80} 
A PDW would naturally be expected to have an anisotropic stiffness if it breaks the rotational symmetry of the crystal. Nevertheless, to explain our observation of the nematic peak in the transverse resistance in terms of fluctuating PDW order would require fine-tuning of the PDW ordering temperature to match the observed $T_{\text{c}}$ very closely. An alternative interpretation, presently being investigated, instead invokes a vestigial PDW state \cite{agterberg2015emergent}, accounting all at once for the nematicity, how this couples to superconducting fluctuations, and the increase of anisotropy as the underdoped critical point is approached \cite{waardh2021nematic}.

\section{Methods}

For film synthesis, we have used atomic-layer-by-layer molecular beam epitaxy (ALL-MBE) \cite{bozovic2001atomic}. LSCO thin films were synthesized on LSAO substrates by sequentially depositing La, Sr and Cu. The atomic fluxes were measured using a quartz crystal monitor and the shuttering times were computer-controlled. The substrate temperature was kept at $650^{\circ}\,$C and the ozone partial pressure at $5 \cdot 10^{-6}$Torr. Reflection high-energy electron diffraction (RHEED) was used to monitor the crystal structure and morphology of the film in real time. The RHEED patterns showed single-crystal LSCO films growing epitaxially on LSAO. The film thickness was controlled digitally by counting the RHEED intensity oscillations. After the growth, the ozone partial pressure was increased to $2\cdot 10^{-5}\,$Torr, the film annealed for 30 minutes (to fill-in oxygen vacancies) and cooled down at the same pressure. 

Subsequently, the films were characterized by \textit{ex-situ} mutual inductance measurements \cite{he2016high}. The real component of inductance showed the Meissner effect with a very sharp onset at $T_{\text{c}}$ indicating remarkable film homogeneity and uniformity. The peak in the imaginary component of inductance, which measures the rise and fall of the ac ($40\,$KHz) conductance, was an order of magnitude sharper than the resistive transition, because the latter is broadened by superconducting fluctuations and vortex flow.

We have verified by high-resolution X-ray diffraction measurements that our LSCO films are almost perfectly tetragonal. The films are very thin and epitaxially anchored to the tetragonal LSAO substrates, so the orthorhombic distortions are suppressed compared to the bulk samples of the same composition. 

To study the angular dependence of longitudinal and transverse resistivity, we patterned the films into devices for transport measurements, using a well-established lithographic process that includes dry etching by ion bombardment. A $500\,$nm thick layer of gold was deposited on the contact pads, ensuring low-resistance Ohmic contacts. A ``sun-beam'' lithography pattern shown in Figure \ref{fig_1}\text{A} consists of 36 current-carrying Hall bars, with the angle $\Delta \phi=10^{\circ}$ between each two consecutive bars. The strips are $100\,\mu$m wide and the voltage contacts spaced $300\,\mu$m part. This pattern enables us to determine the dependence of $\rho$ and $\rho_{\text{T}}$ on the azimuthal angle $\phi$, measured from the $[100]$ crystallographic directions of LSAO and LSCO, with $\pm5^{\circ}$ resolution. 

Transport measurements were made using two experimental setups, based on Helium-4 and Helium-3, and reaching temperatures down to $T = 4.2\,$K and $T=0.3\,$K, respectively. In both setups, the sample is placed in He exchange gas, ensuring the temperature stability better than $\pm1\,$mK. The He-3 setup is equipped with superconducting solenoid magnet capable of reaching the dc field of $9\,$T. The measurements reported here were made with the magnetic field perpendicular to the film surface.

By extensive experimentation that encompassed several thousand devices, we have ruled out every experimental artifact that we could think off, including the possibility that the observed anisotropy in transport originates from film inhomogeneity, misalignment between the pairs of Hall contacts, thermopower due to a thermal gradient, substrate miscut, epitaxial strain gradient, artifacts of lithography, etc. Our large statistics reveals systematic angle, doping, and temperature dependences, rule out random extrinsic factors, and indicate that the observed behavior is intrinsic to LSCO \cite{wu2017spontaneous}. 

\section{Acknowledgements}
Work at Brookhaven National Laboratory was supported by the DOE, Basic Energy Sciences, Materials Sciences and Engineering Division. X. H. is supported by the Gordon and Betty Moore Foundation's EPiQS Initiative through grant GBMF9074.

\textsuperscript{$\ast$ }\textit{Currently at: }Swedish Defense Research Agency (FOI)

\textsuperscript{$\ast \ast$ }\textit{Current address}: Westlake University, Hangzhou, China
\printbibliography
\appendix
%
% change labels in figures
\renewcommand{\thefigure}{S\arabic{figure}}
% set figure counter to 0
\setcounter{figure}{0}    
% change labels in tables
\renewcommand{\thetable}{S\arabic{table}}
% set table counter to 0
\setcounter{table}{0}    
% redefine section label
\renewcommand{\thesection}{S\arabic{section}}
%
% change labels in figures
\renewcommand{\theequation}{S\arabic{equation}}
% set figure counter to 0
\setcounter{equation}{0}   
\clearpage
\newpage
\begin{changemargin}{20pt}{20pt} 
{\LARGE Supplementary Information --- Colossal transverse magnetoresistance due to nematic superconducting phase fluctuations in a copper oxide}
\end{changemargin}
%\section{Supplementary Information}

In this supplemental information we present the details of the modelling outlined and presented under Results and Discussion in the main text. 

\section{\label{sec_S1} Modelling the magneto-resistivity}

\begin{figure}[ht]
\centering
\includegraphics[width=13.84cm,height=8.73cm]{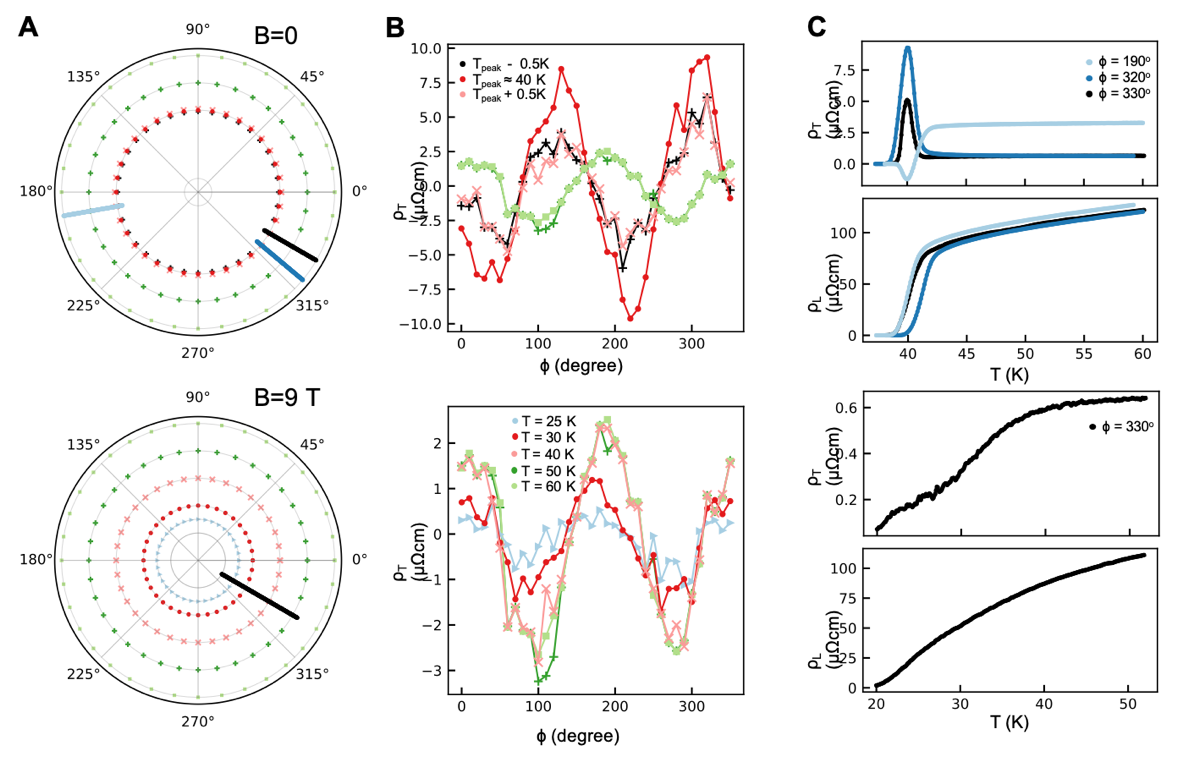}
\caption{\label{fig_S1}\textbf{Complementary presentation of the experimental data of Figure \ref{fig_1} for doping x = 0.16.} The resistivity is measured on all bars for a discrete set of temperatures, as indicated in A. The transverse resistivity is presented in B. The systematic sinusoidal variation of the transverse resistivity is clearly visible both above and near $T_{\text{c}}$, as emphasized further in Figure \ref{fig_S3}, ruling out inhomogeneity as the cause of the peaks. For $B = 0$, the complete temperature dependence of the resistivity is measured for 3 bars, as indicated in A and presented in C. Correspondingly for $B = 9\,$T one bar is studied in detail. (As presented in Figure \ref{fig_1}D and E, also for several magnetic fields in the range $0\,$T to $9\,$T.)}
\end{figure}

In Figure \ref{fig_S1}, the experimental data are presented in a complementary fashion to Figure \ref{fig_1} of the main text.

%
%\subsection{Figure S1| Complementary presentation of the experimental data of Figure1 for doping x = 0.16. The resistivity is measured on all bars for a discrete set of temperatures, as indicated in A. The transverse resistivity is presented in B. The systematic sinusoidal variation of the transverse resistivity is clearly visible both above and near Tc, as emphasized further in Figure S3, ruling out inhomogeneity as the cause of the peaks. For B = 0, the complete temperature dependence of the resistivity is measured for 3 bars, as indicated in A and presented in C. Correspondingly for B = 9 T one bar is studied in detail. (As presented in Figure1 D and E, also for several magnetic fields in the range 0 T to 9 T.)}

Several key features can be observed in the measured data, to guide the modelling. 1) The sharp peaks close to $T_{\text{c}}$ are evidently intimately tied to superconductivity. We model these as arising from superconducting fluctuations. 2) The peaks show a clear nematic signature. Thus, fluctuations need to have an in-plane anisotropy. 3) Weaker nematicity survives up to high temperatures, not likely attributable to superconducting fluctuations. We model this as conductivity from normal electrons with a finite mass anisotropy. We thus consider a two-fluid model of conductivity consisting of normal electrons and thermally excited Cooper-pairs. 5) The nematic director identified from the nodes of the transverse resistivity depends on temperature. To model this, we will allow for the normal and superconducting components to have different principal axes orientations. To simplify the modeling, we will assume that both components have a temperature-independent orientation. Then, any net rotation will solely arise from a redistribution of the relative weight of the two channels. Below we describe these two conducting channels. 

\subsection{Anisotropic paraconductivity}

As explained in the main text, our rudimentary model for the magnetoresistance consists of normal electrons described by an anisotropic Drude model and damped anisotropic Cooper-pairs. The contribution from the latter to the conductivity (\eqref{eq_6} in the main text) generalizes the derivation presented in Ref. \cite{larkin2005theory} based on Aslamazov and Larkin \cite{aslamazov1968the} to the case of an anisotropic mass tensor. 

The basics of the paraconductivity is that even though Cooper-pairs are not stable and cannot provide a dissipation-free supercurrent above $T_{\text{c}}$, they are still available as charge-carrying excitations with their own dynamics. We can describe this with a time-dependent Ginzburg Landau (TDGL) equation, which we use to study the weak-field response to an electric field $\vec{E}  = -\triangledown \varphi -\frac{\partial\vec{A} }{\partial t}$. Cast as a Langevin equation, the TDGL equation takes the form:
\begin{equation}
\left[\hat{L}^{-1}+i2e\gamma \hat{\varphi}\right]\psi  = \zeta
\label{TDGL}
\end{equation}
where the thermal fluctuations are described by the inclusion of a Langevin force $\zeta$ whose correlations are given by:

\begin{equation}
\langle \zeta^{\ast }(r,t)\zeta (r',t')\rangle  = 2T \text{Re}(\gamma)\hbar \delta (t-t')\delta (r-r')\,.
\end{equation}

The operator $\hat{L}$ describes the propagation of pairs:
\begin{eqnarray}
%\begin{matrix}
 \hat{L} &  = & \left(\gamma \hbar\frac{\partial }{\partial t}+\right)^{-1}, \\ 
 \hat{H}(\hat{k}_{i}-\frac{2eA_{i}}{\hbar }) & = & a\left(\epsilon +\xi_{0,i}^{2}(_{i}-\frac{2eA_{i}}{\hbar })^{2}\right) 
%\end{matrix}
\end{eqnarray}
where $\epsilon  =\frac{T-T_{\text{c}}}{T_{\text{c}}}$ and $\xi_{0,i}^{2} =\frac{\hbar^{2}}{2m_{\text{p},i}a}$ is the anisotropic coherence length, attributed to the anisotropic pair-mass $m_{\text{p},i}$, where $i,j$ refer to the principal-axes frame, and $a$ has the units of energy. The dynamics is determined by the coefficient $\gamma$, which is complex in general. Assuming a completely relaxational dynamics, we take $\gamma$ as real in what follows; this is justified in the weak-coupling BCS limit where the dissipation is dominated by the breaking of Cooper-pairs. The equilibrium solution of \eqref{TDGL} in the absence of an electric field is given by $\psi^{(0)} = \zeta$. We can solve the full equation to the first order in $\varphi$ (we choose a gauge where $\hat{A} = 0$) by writing $\psi  = \psi^{(0)}+\psi^{(1)}$, where $\psi^{(1)}\propto \varphi$, and expanding in $\varphi$. We find $\psi^{(1)} = -i2e\gamma\hat{L}\hat{\varphi}\hat{L} \zeta$. The homogeneous current takes the form:
\begin{equation}
J_{i}(t) =\frac{1}{V}\frac{4e\hbar }{m_{B}^{i}}\text{Re} Tr\left(\hat{k}_{i}\vert \psi^{(1)}(t)\rangle \langle \psi^{(0)}(t)\vert\right)\,.
\end{equation}

Inserting the solutions found above together with the scalar potential, $\hat{\varphi} = -E_{j}\hat{r}_{j}$ for a constant electric field $ E_{i}$, and averaging over the Langevin forces we find:
\begin{equation}
%\begin{matrix}
J_{i}  = -E_{j}\frac{1}{V}\frac{16T\gamma^{2}\hbar^{2}e^{2}}{m_{B}^{i}}\times  
  \text{Im} \int\frac{d\omega }{2\pi }Tr\left(\hat{k}_{i}\hat{L}(\omega)\hat{r}_{j}\hat{L}(\omega)\hat{L}^{†}(\omega)\right). 
%\end{matrix}
\end{equation}

The trace can be evaluated in the momentum representation where the propagator $ L(k,\omega) =\frac{1}{-i\gamma \hbar \omega +H(k)}$ is diagonal. Using the canonical commutation relations, the position operator can be written as:
\begin{equation}
\langle k\vert\hat{r}_{j}\vert k'\rangle  = -i\frac{\hbar^{2}p_{j}}{m_{B}^{j}}\frac{(2\pi)^{2}\delta (k-k')}{H(k)-H(k')}
\end{equation}
from which we can derive the conductivity after integrating over frequencies:
\begin{equation}
%\begin{matrix}
\sigma_{ij}^{2D} =\frac{2T\gamma e^{2}\hbar^{3}}{m_{B}^{j}m_{B}^{i}}\int\frac{d^{2}k}{(2\pi)^{2}}\frac{k_{i}k_{j}}{H(k)^{3}}  
%\end{matrix}
\end{equation}
which is evidently diagonal. Given the definition of $ H(k)$ and the fact that 
$\xi_{k}^{2} =\frac{\hbar^{2}}{2m_{\text{p},i}a\epsilon }$, we find:
\begin{eqnarray}
%\begin{matrix}
\sigma_{ii}^{2D} &  =  & \frac{2T\gamma e^{2}(\xi_{i}^{2})^{2}}{\pi^{2}(a\epsilon)\hbar }\int dk_{x}dk_{y}\frac{k_{i}^{2}}{(1+\xi_{k}^{2}k_{k}^{2})} \nonumber\\ 
 &  = & \frac{T\gamma e^{2}}{2\pi \hbar a\epsilon }\frac{\xi_{i}^{2}}{\xi_{x}\xi_{y}}\,.
%\end{matrix}
\end{eqnarray}

Using the BCS expression $\gamma  =\frac{\pi a}{8T}$ we arrive at: 
\begin{equation}
\sigma_{ii}^{2D} =\frac{e^{2}}{16\hbar \epsilon }\frac{\xi_{i}^{2}}{\xi_{x}\xi_{y}} =\frac{e^{2}}{16\hbar \epsilon }\sqrt{\frac{m_{\text{p},x}m_{\text{p},y}}{(m_{\text{p},i})^{2}}} 
\end{equation}
which, after division by the layer thickness $d$, and rewriting as a tensor provides the expression \eqref{eq_6} in the main text.

\subsection{Conductivity in magnetic field}

When a field is applied, an antisymmetric Hall part of the conductivity tensor, $\sigma_{\text{H}}$, is induced, yielding an antisymmetric term in the resistivity. This is measured and subtracted from the data presented in Figure \ref{fig_1} and Figure \ref{fig_S1}. The magnetic field also acts to increase the resistivity, the so-called magneto-resistivity, which is addressed below. 

\subsubsection{Drude conductivity in magnetic field}

In the principal-axes frame, the conductivity in the Drude channel takes the form: 
\begin{equation}
\overline{\sigma }_{\text{n}}\left(B\right) =\frac{\sigma_{\text{D}}}{1+(\omega_{\text{c}}\tau)^{2}}\begin{bmatrix}
\lambda_{\text{n}} & \omega_{\text{c}}\tau  \\ 
-\omega_{\text{c}}\tau  &\frac{1}{\lambda_{\text{n}}} \\ 
\end{bmatrix}.
\end{equation}

where $\sigma_{\text{D}} =\frac{nq^{2}\tau }{\sqrt{m_{x}^{\ast }m_{y}^{\ast }}}$ gives the Drude expression in zero field, $\lambda_{\text{n}}  =\sqrt{m_{\text{n},y}/m_{\text{n},x}}$ is the mass anisotropy, $\tau $  is the relaxation time, and $\omega_{\text{c}} =\frac{qB}{\sqrt{m_{x}^{\ast }m_{y}^{\ast }}}$ is the cyclotron frequency. The Hall conductivity depends on $\omega_{\text{c}}\tau$ to first order, while the magnetoresistance depends on $\omega_{\text{c}}\tau$ to second order. Since $B \leq 9\,$T in our measurements, we expect $\omega_{\text{c}}\tau$ to be small; e.g., from the Hall-coefficient measurements in Ref.\ \cite{murayama1991correlation} we estimate $\omega_{\text{c}}\tau \approx  0.01$ right above $ T_{\text{c}}$ for $B = 9\,$T and $x=0.16$. The Hall conductivity is subtracted from data, and since $\omega_{\text{c}}\tau$ is small, we neglect the influence of a magnetic field on the Drude conductivity altogether.

Note that this is an approximation, which can indeed cause some deviation from the experiment. The actual field dependence of magnetoresistance is in fact that of a ``strange-metal'' rather than that of a Fermi Liquid \cite{giraldo2018scale}. But since the underlying cause is not well understood, and to keep the modelling simple, we are not including that here.

\subsubsection{Paraconductivity in magnetic field}

In contrast to the normal channel, the superconducting channel, $\sigma_s$, is strongly field dependent. The effect is primarily made up of two parts. The first is related to a change in the paraconductivity, whose main effect, as follows, is to suppress $T_{\text{c}}$. In the presence of a magnetic field the fluctuating contribution to conductivity takes the form \cite{larkin2005theory}:
\begin{equation}
\overline{\sigma }_{s}\left(B\right) =\begin{bmatrix}
\sigma_{sx}\left(B\right) & \sigma_{s}^{H}\left(B\right) \\ 
-\sigma_{s}^{H}\left(B\right) & \sigma_{sy}\left(B\right) \\ 
\end{bmatrix},
\end{equation}
with
\begin{equation}
\begin{matrix}
\sigma_{s(x,y)}(B) &  =\frac{e^{2}}{2\hbar }F\left(\frac{\epsilon }{2b}\right)\frac{(\lambda,\lambda^{-1})}{\epsilon }\,,  
\end{matrix}
\end{equation}
where $\epsilon  = \ln (T/T_{\text{c}})\approx\frac{T_{\text{c}}}{T-T_{\text{c}}}$, $ b = B/B_{\text{c}2}$, and $ B_{\text{c}2} =\frac{\Phi_{0}}{2\pi \xi_{x}(0)\xi_{y}(0)}$ is the (zero-temperature) upper critical field. Here, we have introduced the function:
\begin{equation}
\begin{matrix}
F(x) &  = x^{2}\left(\psi (x+1/2)-\psi (x)-\frac{1}{2x}\right) \\ 
\end{matrix}
\end{equation}
where $\psi (x)$ is the digamma function. The most important effect is the change of the transition temperature, which will be given by the first singularity of $ F(x)$ situated at $ x = -1/2$, which yields $ T_{\text{c}}(H) = T_{\text{c}}(0)e^{-b}$. The fluctuation contribution to the Hall conductivity, $\sigma_{\text{p}}^{H}$, is proportional to $\text{Im}(\gamma)$, which in BCS is expected to be small ($\text{Im}\left(\gamma\right)\sim\frac{T_{\text{c}}}{T_{F}}\text{Re}(\gamma)$). 

\subsubsection{Broadening due to flux motion}

\begin{figure}[ht]
\centering
\includegraphics[width=0.5\textwidth]{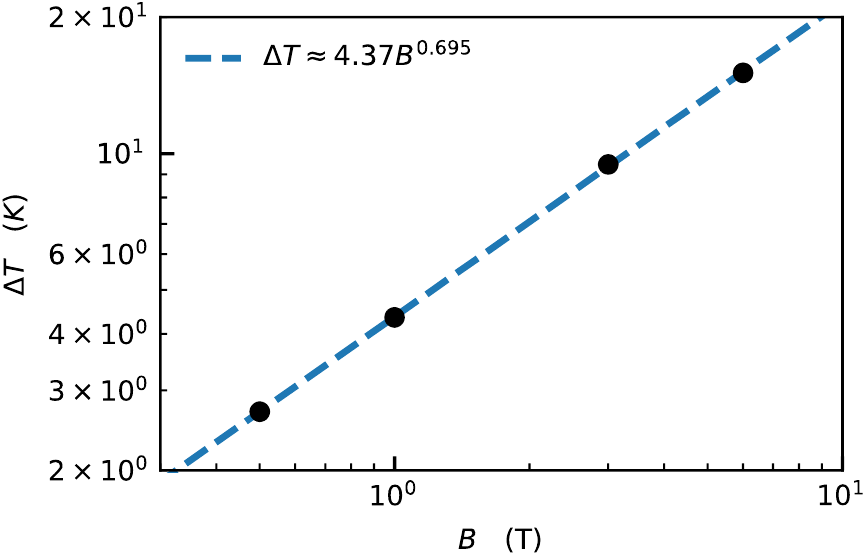}
\caption{\label{fig_S2}\textbf{Resistive broadening} The transverse resistivity, $\rho_{\text{T}}(T)$, measured in the same direction. The peak in $\rho_{\text{T}}(T)$ is suppressed by the magnetic field, indicating that it is of a superconducting origin. Fitting the resistive broadening to $\Delta T\propto B^{2/3}$ (see Ref. \cite{tinkham1988resistive}) where $\Delta T$ is defined as the half-width of $d\rho_{\text{T}}/dT$ near $T_{\text{c}}$. The observed exponent is $\gamma =$ 0.695.}
\label{fig:transverse_resistivity_tt_measured_same}
\end{figure}

The second effect of the magnetic field on the paraconductivity is the broadening of the resistive transition due to flux motion. We use the qualitative result \cite{tinkham1988resistive} that the width of the transition is given by $\Delta T\propto B^{2/3}$, a relation which is in good qualitative agreement with our experiments as can be seen in Figure \ref{fig_S2}. To model the conductivity tensor, we therefore include this broadening 
\begin{equation}
   \Delta T_{\text{c}}\left(b\right) = \Delta T_{\text{c}}\left(0\right)+\beta b^{\frac{2}{3}} 
   \label{eq:broadening}
\end{equation} 
$ b = B/B_{\text{c}2}$ in the Gaussian average over $T_{\text{c}}$, where $\Delta T_{\text{c}}\left(0\right)$ corresponds to the broadening used for the $B=0$ fit. 

\begin{table}[ht]
\footnotesize
%\begin{adjustbox}%{max width=\textwidth}
\begin{center}
\begin{tabular}{p{1.13cm}p{3.47cm}p{3.97cm}p{0.81cm}}
\hline
\multicolumn{1}{|p{1.12cm}}{{ x}} & 
\multicolumn{1}{|p{3.45cm}}{$\alpha(T_{\text{c}})$ (degrees)} & 
\multicolumn{1}{|p{3.96cm}}{$\alpha(295K)$ (degrees)} & 
\multicolumn{1}{|p{0.83cm}|}{$\nu$} \\ 
\hline
\multicolumn{1}{|p{1.12cm}}{{ 0.04}} & 
\multicolumn{1}{|p{3.45cm}}{60} & 
\multicolumn{1}{|p{3.96cm}}{62} & 
\multicolumn{1}{|p{0.83cm}|}{2} \\ 
\hline
\multicolumn{1}{|p{1.12cm}}{{ 0.10}} & 
\multicolumn{1}{|p{3.45cm}}{24} & 
\multicolumn{1}{|p{3.96cm}}{16} & 
\multicolumn{1}{|p{0.83cm}|}{-8} \\ 
\hline
\multicolumn{1}{|p{1.12cm}}{{ 0.12}} & 
\multicolumn{1}{|p{3.45cm}}{163} & 
\multicolumn{1}{|p{3.96cm}}{179} & 
\multicolumn{1}{|p{0.83cm}|}{16} \\ 
\hline
\multicolumn{1}{|p{1.12cm}}{{ 0.16}} & 
\multicolumn{1}{|p{3.45cm}}{81} & 
\multicolumn{1}{|p{3.96cm}}{142} & 
\multicolumn{1}{|p{0.83cm}|}{61} \\ 
\hline
\multicolumn{1}{|p{1.12cm}}{{ 0.18}} & 
\multicolumn{1}{|p{3.45cm}}{113} & 
\multicolumn{1}{|p{3.96cm}}{132} & 
\multicolumn{1}{|p{0.83cm}|}{19} \\ 
\hline
\multicolumn{1}{|p{1.12cm}}{{ 0.21}} & 
\multicolumn{1}{|p{3.45cm}}{120} & 
\multicolumn{1}{|p{3.96cm}}{120} & 
\multicolumn{1}{|p{0.83cm}|}{0} \\ 
\hline
\end{tabular}
\end{center}
%\end{adjustbox}
\caption{\label{tab_S1}\textbf{The angle of resistivity principal axis at $\mathbf{T=T_{\text{c}}}$ and at $\mathbf{T=295}\,$K.} These are obtained from fitting the angular dependence of the transverse response to $\rho_{\text{T}}\sin (2(\phi-\alpha))$ with $\nu=\alpha(295K)-\alpha(T_{\text{c}})$, see Figure \ref{fig_S3}.}
\label{tab:table_angle_resistivity_principal_axis}
\end{table}

\begin{figure}[ht]
\centering
\includegraphics[width=0.9\textwidth]{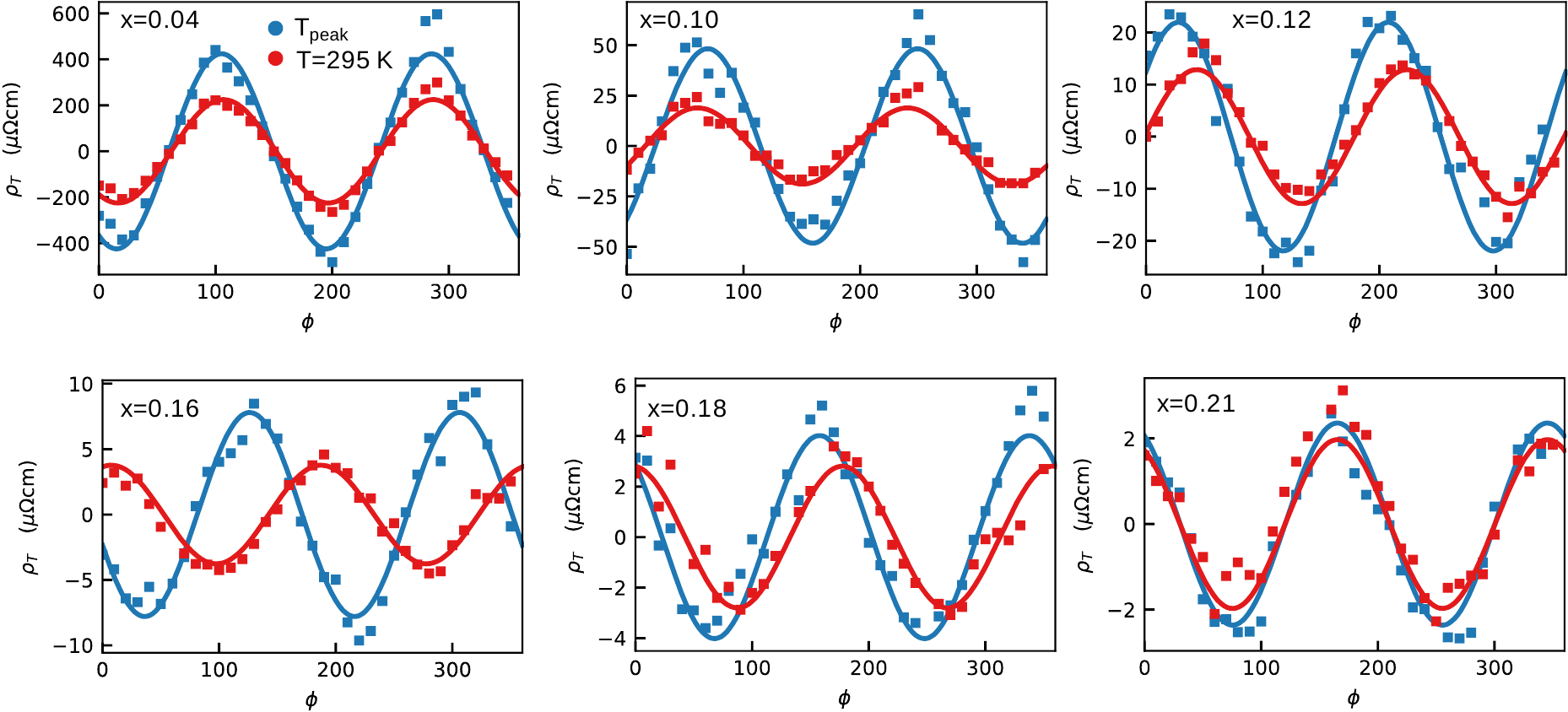}
\caption{\label{fig_S3}\textbf{Fit of the angular dependence of the transverse response for different doping.} The red and blue squares correspond to measured data at $T_{\text{c}}$ and at $T= 295\,$K, respectively. The solid lines indicate the best fit to $\rho_{\text{T}}\sin (2(\phi-\alpha))$, with $\alpha$ listed in Table \ref{tab_S1}. The error of the fits are presented in Table \ref{tab_S4}.}
\label{fig:_fit_angular_dependence_transverse}
\end{figure}
%
%Note that the amplitude at $T_{\text{c}}$ is higher than at $T=295$K, consistent with a relatively strong anisotropy for all doping levels.
%

\section{Modeling the nematic state (Nematicity and conductivity in tetragonal symmetry)}

In a tetragonal crystal there are two distinct symmetry-allowed nematic fields: a bond-nematic, ($B_{1g}$) field $N_{1g}$ transforming as $x^2-y^2$, and a diagonal-nematic, ($B_{2g}$) field $N_{2g}$ transforming as $xy$. It is important to note that these two fields transform independently, and one should take care when considering the corresponding matrix
\begin{equation}
\overline{N} =\begin{bmatrix}
N_{1g} & N_{2g} \\ 
N_{2g} & -N_{1g} \\ 
\end{bmatrix}
\end{equation}
since it is not a proper order-parameter. (In group-theory terms, $\overline{N}$ is reducible.) This is in contrast to a continuous 2D rotational symmetry, where the symmetry breaking is categorized by a nematic director. In the tetragonal system, no such unique nematic director exists. Specifically, the angle $\alpha_{N} =\frac{1}{2}\arctan\left(N_{2g}/N_{1g}\right)$ is not a unique classifier of the symmetry breaking state. (Except for  $\alpha_{N} = 0$ or $\alpha_{N} = \pi/4$ modulo $\pi/2$, which are distinctive indicators of pure $B_{1g}$ or $B_{2g}$ symmetry, respectively.) 

\subsection{Nematicity and conductivity}

In the main text we saw that in the absence of a symmetry breaking magnetic field ($\sigma_{\text{H}} = 0$) the development of a transverse resistivity is equivalent to a finite traceless symmetric part of the conductivity matrix  
\begin{equation}
\overline{\sigma }_{S} =\begin{bmatrix}
\sigma_{S}^{(1)} & \sigma_{S}^{(2)} \\ 
\sigma_{S}^{(2)} & -\sigma_{S}^{(1)} \\ 
\end{bmatrix}
\end{equation}
with the longitudinal and transverse conductivities given by
\begin{eqnarray}
\rho_{\text{T}}(\phi) &  = &\frac{\sigma_{\text{H}}+\sqrt{\vert det\left(\overline{\sigma }_{S}\right)\vert }\sin (2(\phi -\alpha))}{\sigma_{\text{H}}^{2}+\sigma_{0}^{2}+\det\left(\overline{\sigma }_{S}\right)} \\ 
\rho_{\text{L}}(\phi) &  = &\frac{\sigma_{0}+\sqrt{\vert det\left(\overline{\sigma }_{S}\right)\vert }\cos (2(\phi -\alpha))}{\sigma_{\text{H}}^{2}+\sigma_{0}^{2}+\det\left(\overline{\sigma }_{S}\right)}\,. 
\end{eqnarray}
Here, $\alpha  =\frac{1}{2}\arctan\left(\sigma_{S}^{(2)}/\sigma_{S}^{(1)}\right)$ is the angle of the principal axis of the conductivity, measured from the crystal $a$-axis. This form of the transverse resistivity agrees well with the measured samples, see Figure \ref{fig_S3}, from which the director $\alpha$ can be extracted (see Table \ref{tab_S1}). 

The previous discussion of the nematic field can directly be carried over to the conductivity where $\sigma_{S}^{(1)}$, $\sigma_{S}^{(2)}$ can be considered nematic order-parameters. Alternatively, we can consider $ N_{1g}$ and $ N_{2g}$ as the fundamental nematic fields that couple to the conductivity $  \sigma_{S}^{(2)} = \gamma_{2}N_{2g}$, $\sigma_{S}^{(1)} = \gamma_{1}N_{1g}$, where $\gamma_{1},  \gamma_{2}$ are independent. 

Thus, we attribute the finite transverse resistivity to the development of nematic order. Furthermore, from the extracted values of the director $\alpha$ in Table \ref{tab_S1} and Figure \ref{fig_S3} we can directly infer the presence of both\textit{B}\textsubscript{1g} and \textit{B}\textsubscript{2g }order-parameters, consistent with a broken tetragonal symmetry.

One problem with directly associating the traceless symmetric part of the conductivity tensor, $   \sigma_{S}^{(2)}$ and $\sigma_{S}^{(1)}$, with the nematic order-parameter, is that it leads to a very complicated behavior of the nematic state as a function of temperature. As can easily be seen from Figure \ref{fig_S1},\ref{fig_S2} and Table \ref{tab_S1}, the anisotropic part of the conductivity has a strong temperature dependence. Instead, it turns out that associating the nematicity to the Drude and paraconductivity mass tensors yields a very simple model, which is in good agreement with the data.

\section{\label{sec_S3} Two-fluid model of conductivity}

The transverse resistivity has a very pronounced peak near $T_{\text{c}}$, as can easily be seen from Figure \ref{fig_S1}C. Instead of associating this with an abrupt change in the nematic order, we will show that this can be attributed to the rise in superconducting fluctuations, within an approximately temperature independent nematic order which is quantified by the anisotropy of the mass tensors.

Another seemingly peculiar property of the nematic state as a function of temperature is the apparent twist in the nematic state as a function of temperature (best visualized in Figure \ref{fig_1} in the main text but inferable from Table \ref{tab_S1} and Figure \ref{fig_S3}). However, as it was discussed above, the angle $\alpha$ should not be mistaken for a nematic director, and consequently a change in $\alpha$ is not necessarily indicative of any change in the underlying nematic order. The simplest model is again to assert a static, temperature independent, nematic order, categorized by $ N_{1g}$ and $ N_{2g}$. However, in order to account for the twist in $\alpha  =\frac{1}{2}\arctan\left(\sigma_{S}^{(2)}/\sigma_{S}^{(1)}\right)$ we will assume that the nematic order-parameters couple to the two conducting channels independently. The twist in $\alpha$ within this model will solely be due to the nature of the different conductivity components. One fluid will be constituted by fluctuations of Cooper-pairs, described by paraconductivity, and the other a normal Drude-like current. We find that such a static nematic state can be realized by associating the nematic fields $ N_{1g}$ and $ N_{2g}$ to the traceless symmetric parts of the normal and superconducting mass tensors, $\overline{m}_{n}$and $\overline{m}_{\text{p}}$, respectively.

Just as for the traceless symmetric part of the conductivity, the traceless symmetric part of the mass tensors can be coupled to the nematic fields. Specifically for the superconducting mass we write 

\begin{equation}
(1/\overline{m}_{\text{p}})_{ij} =\left(\frac{1}{2m}\right)_{ij}+ S_{ij}\,,
\end{equation}
where $S$ is a rank-two traceless (reducible) symmetric tensor with $ S_{xx} = -S_{yy} =\gamma_{SC-1}N_{1g}$, $ S_{xy} = S_{yx} =\gamma_{SC-2}N_{2g}$, and $m$ the normal isotropic electron mass. Similarly, for the single-particle (unpaired) electrons 
\begin{equation}
(1/\overline{m}_{n})_{ij} =\left(\frac{1}{m}\right)_{ij}+ S'_{ij}\,,
\end{equation}
where $S’$ is another tensor made up by the same nematic fields as above, but with distinct coupling constants $\gamma_{N-1},\gamma_{N-2}$. Here we see that our two-fluid model will carry the two distinct principal axes of $\overline{S}$ and $\overline{S}'$, given by $\alpha_{S} =\frac{1}{2}\arctan\left(S_{xy}/S_{xx}\right)$, $\alpha'_{S} =\frac{1}{2}\arctan\left(S'_{xy}/S'_{xx}\right)$, respectively. This model provides a natural explanation for the twist of the principal axes of $\overline{\sigma }$ as a result of the change with temperature in the relative weight of the $\overline{\sigma }_{n}$ and $\overline{\sigma }_{\text{p}}$ contributions. At high temperatures, $\overline{\sigma }_{n}$ dominate and we find $\alpha \approx \alpha'_{S}$, while at temperatures just above $T_{\text{c}}$ the principal axis is given by $\alpha \approx \alpha_{S}$. This also naturally explains the confinement of the twist to the temperatures range just above $T_{\text{c}}$, in which superconducting fluctuations are dominant. 

\subsection{Temperature independent nematic order, modelled by mass tensors}

The precise evolution of the conductivity $\overline{\sigma }$ as a function of temperature and doping will depend on the parameters  $\gamma_{N-1},\gamma_{N-2},\gamma_{SC-1},\gamma_{SC-2}$ and the fields $N_{1g}$, $N_{2g}$, which all in principle depend on both temperature and doping. However, as already mentioned, we find that the measurements presented here, as well as those in Ref. \cite{wu2017spontaneous}, can be well accounted for by assuming that both the nematic order, $N_{1g}$, $N_{2g}$ and the couplings that generate the mass anisotropies, are independent on temperature. (Or equivalently, the traceless symmetric parts of the mass tensors are independent on temperature.) This is important; it indicates that the energy scale of the nematic order is large enough for it to be well-developed already at room temperature. Consequently, we model the temperature dependence only through the fixed form given by the expression (\eqref{eq_6} for the paraconductivity and by the temperature dependence of the Drude relaxation time.

Assuming a temperature independent nematic state also implies that we can model the conductivity phenomenologically at the level of $\overline{m}_{n}$ and $\overline{m}_{\text{p}}$. This will give information about the masses that is independent of the microscopic mechanism of nematicity.

\subsection{Twist of principal axes of conductivity}

To account for the twist, we align $\overline{m}_{n}$  and $\overline{m}_{\text{p}}$ with the measured $\rho_{\text{T}}$ at the high-temperature tail, and just above $ T_{\text{c}}$, respectively. The measured principal direction of $\overline{\sigma }$ for these temperatures are presented in Table \ref{tab_S1}  and the related Figure \ref{fig_S3}. We see that the twist is  most pronounced for the $x=0.16$ sample, while the other samples show a much smaller twist. To a good approximation, we will therefore align the principal axis of $\overline{m}_{n}$ and $\overline{m}_{\text{p}}$  with one another for all samples except for $x=0.16$.

\subsection{\label{sec_S3.3} Basic model}

Before embarking on a more complete fit, we here focus on the most basic features of the two-fluid model in the absence of magnetic field. The objective is to show that even a very simple model can explain our main experimental findings, including the difference in the anisotropy of the normal and superconducting components that the data imply. 

Assume that the principal axis of $\overline{\sigma }_{n}$ and $\overline{\sigma }_{\text{p}}$ are aligned. In this principal frame $(e_x,e_y)$, we find the full conductivity as

\begin{equation}
\overline{\sigma } =\overline{\sigma }_{n}+\overline{\sigma }_{\text{p}} = \begin{bmatrix}
\sigma_{n,x}+\sigma_{\text{p},x} & 0 \\ 
0 & \sigma_{n,y}+\sigma_{\text{p},y} \\ 
\end{bmatrix}
\label{eq:sig_simple}
\end{equation}
Here the paraconductivity is given by 

\begin{equation}
\begin{matrix}
\sigma_{\text{p},x} &  = \sigma_{\text{p},0}\frac{T_{\text{c}}}{T-T_{\text{c}}}\lambda  \\ 
\sigma_{\text{p},y} &  = \sigma_{\text{p},0}\frac{T_{\text{c}}}{T-T_{\text{c}}}\lambda^{-1} \\ 
\end{matrix}
\end{equation}

where $\sigma_{\text{p},0} =\frac{e^{2}}{16\hbar d}$ and $\lambda =\sqrt{m_{\text{p},y}/m_{\text{p},x}}$ parametrizes the anisotropy of mass of Cooper-pairs. Here we will treat $\sigma_{\text{p},0}$  as a free parameter and in the next section explore how well it agrees with the theoretical value while fitting the model to the data. The normal component is given by the Drude form $\sigma_{n,x} =  \lambda_{n} \sigma_{\text{D}}$, $\sigma_{n,y} =  \lambda_{n}^{-1} \sigma_{\text{D}}$,  with $\lambda_{n}  =\sqrt{m_{\text{n},y}/m_{\text{n},x}}$ and $\sigma_{\text{D}} =\frac{ne^{2}\tau }{\sqrt{m_{\text{n},y}m_{\text{n},x}}}$. The full longitudinal and transverse conductivities are given by
\begin{eqnarray}
\rho_{\text{T}}(\phi) &  = &\rho_{\text{T}}\sin (2(\phi -\alpha)) \\ 
\rho_{\text{L}}(\phi) &  = & \rho_{\text{L}}+\rho_{\text{T}}\cos\left(2\left(\phi -\alpha\right)\right)\,, 
\label{eq:rho_simple}
\end{eqnarray}
where $\rho_{\text{L}} =\frac{\rho_{x}+\rho_{y}}{2} = (\sigma_{x}+\sigma_{y})/2\sigma_{x}\sigma_{y}$ and $\rho_{\text{T}} =\frac{\rho_{x}-\rho_{y}}{2} = (\sigma_{y}-\sigma_{x})/2\sigma_{x}\sigma_{y}$ while $\alpha$ is the angle between $e_x$ and the crystal $a$-axis, i.e., the $[100]$ direction. 

\begin{figure}[ht]
\centering
\includegraphics[width=0.9\textwidth]{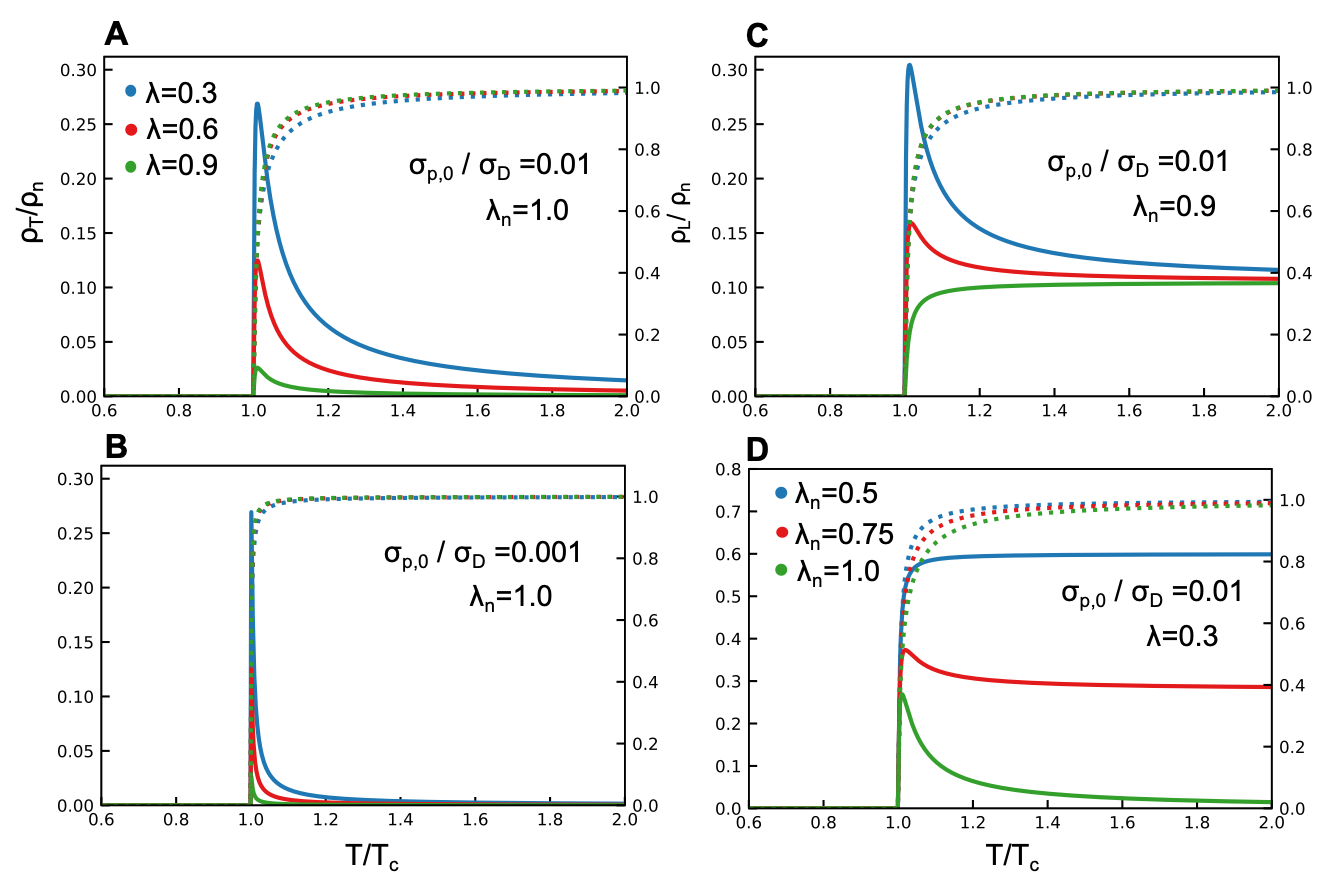}
\caption{\label{fig_S4}\textbf{Basic model.} The transverse (solid) and longitudinal (dashed) resistivity predicted from \eqref{eq:sig_simple}-\ref{eq:rho_simple}, evaluated at $(\phi-\alpha)=\pi/4$, assuming a constant normal resistivity component and anisotropic superconducting fluctuations. In the panels A and B three different superconducting mass anisotropies $\lambda$ are shown for two different values of $\sigma_{\text{p},0}/\sigma_{\text{D}}=$ 0.01 and 0.001, respectively, while setting the normal component anisotropy to zero (i.e., $\lambda_{\text{n}}=$1). In the panel C we instead set $\lambda_{\text{n}}=0.9$ for the same values of $\lambda$ and here we see a finite asymptotic value for high temperatures. In the panel D we vary $\lambda_{\text{n}}$ for fixed $\sigma_{\text{p},0}/\sigma_{\text{D}}$ and $\lambda$. By decreasing $\lambda_{\text{n}}$, the peak in the transverse response vanishes. Without superconducting anisotropy, $\lambda=1$, we find no peak whatsoever.}
\label{fig:model_transverse_solid_and_longitudinal}
\end{figure}

A few examples of the temperature dependence of $\rho_{\text{T}}$ and $\rho_{\text{L}}$ from this model are presented in Figure \ref{fig_S4}, assuming a $T$-independent normal component $\sigma_{\text{D}}\left(T\right) = \sigma_{\text{D},0}$, in the relevant regime $\sigma_{\text{p},0}\ll  \sigma_{\text{D},0}$. (We take $\sigma_{y}>\sigma_{x}$ such that  $\lambda,\lambda_{n}\leq 1$.) The main features of this model can be summarized as follows. 
1) The contribution to the \textit{longitudinal} resistivity from the paraconductivity is small except very close to $ T_{\text{c}}$, where it causes a rapid drop in resistivity. Further, $\rho_{\text{L}}$ is insensitive to both mass anisotropies $\lambda $ and $\lambda_{n}$. 2) For the \textit{transverse} resistivity, which is sensitive to the difference between $\sigma_{x}$ and $\sigma_{y}$, the situation is dramatically different. With a finite superconducting anisotropy, $\lambda <1$, the response is peaked near $ T_{\text{c}}$, reflecting the anisotropy of the approach to the singularity of paraconductivity at $ T_{\text{c}}$. A normal component anisotropy, $\lambda_{n}<1,$ enhances the high-temperature tail of the transverse response but suppresses the peak. In fact, the existence of a peak requires $  \lambda <\lambda_{n}^{2}$. Thus, superconducting anisotropy is necessary to have a peak, and it must be bigger than the anisotropy of the normal component\footnote{For $\lambda>1$ the requirement for a peak reads $\lambda>\lambda_{\text{n}}^2$}.

This simple model is in a good qualitative agreement with the zero-field data (see Figure \ref{fig_S1} and Figure \ref{fig_S6}). It predicts a well-pronounced peak and a residual tail of the transverse resistivity, provided $\lambda \ll \lambda_{n}$, i.e., if the pair-mass anisotropy is significantly larger than the normal state anisotropy.

\begin{table}[ht]
\footnotesize
%\begin{adjustbox}{max width=\textwidth}
\begin{center}
\begin{tabular}{p{1.25cm}p{2.16cm}p{2.0cm}p{2.63cm}p{2.97cm}p{2.09cm}p{2.09cm}}
\hline
\multicolumn{1}{|c|}{} &
\multicolumn{2}{|c|}{Peak} & 
\multicolumn{2}{|c|}{Linear} & 
\multicolumn{1}{|c|}{Quotient} & 
\multicolumn{1}{|c|}{Extended Fit} \\ 
\hline
\multicolumn{1}{|c|}{x} & 
\multicolumn{1}{|c|}{$ m_{\text{p},y}/m_{\text{p},x}$} & 
\multicolumn{1}{|c|}{$\sigma_{\text{p},0} (\frac{e^{2}}{16\hbar d})$} & 
\multicolumn{1}{|c|}{$ m_{\text{p},y}/m_{\text{p},x}$} & 
\multicolumn{1}{|c|}{$\sigma_{\text{p},0} (\frac{e^{2}}{16\hbar d})$} & 
\multicolumn{1}{|c|}{$ m_{\text{p},y}/m_{\text{p},x}$} & 
\multicolumn{1}{|c|}{$ m_{\text{p},y}/m_{\text{p},x}$} \\ 
%x & $ m_{\text{p},y}/m_{\text{p},x}$ & $\sigma_{\text{p},0} (\frac{e^{2}}{16\hbar d})$ & $ m_{\text{p},y}/m_{\text{p},x}$ & $\sigma_{\text{p},0} (\frac{e^{2}}{16\hbar d})$ & $ m_{\text{p},y}/m_{\text{p},x}$ & $ m_{\text{p},y}/m_{\text{p},x}$} \\ 
\hline
\multicolumn{1}{|c|}{0.04} & 
\multicolumn{1}{|c|}{-} & 
\multicolumn{1}{|c|}{-} & 
\multicolumn{1}{|c|}{0.25} & 
\multicolumn{1}{|c|}{2.96} & 
\multicolumn{1}{|c|}{0.31} & 
\multicolumn{1}{|c|}{0.207$\pm$0.04} \\ 
\hline
\multicolumn{1}{|c|}{0.10} & 
\multicolumn{1}{|c|}{0.06} & 
\multicolumn{1}{|c|}{2.23} & 
\multicolumn{1}{|c|}{0.064} & 
\multicolumn{1}{|c|}{11.9} & 
\multicolumn{1}{|c|}{0.024} & 
\multicolumn{1}{|c|}{0.128$\pm$0.11} \\ 
\hline
\multicolumn{1}{|c|}{0.12} & 
\multicolumn{1}{|c|}{0.32} & 
\multicolumn{1}{|c|}{1.49} & 
\multicolumn{1}{|c|}{0.52} & 
\multicolumn{1}{|c|}{52.7} & 
\multicolumn{1}{|c|}{0.49} & 
\multicolumn{1}{|c|}{0.553$\pm$0.06} \\ 
\hline
\multicolumn{1}{|c|}{0.16} & 
\multicolumn{1}{|c|}{0.47} & 
\multicolumn{1}{|c|}{1.59} & 
\multicolumn{1}{|c|}{0.33} & 
\multicolumn{1}{|c|}{7.0} & 
\multicolumn{1}{|c|}{-} & 
\multicolumn{1}{|c|}{0.765$\pm$0.20} \\ 
\hline
\multicolumn{1}{|c|}{0.18} & 
\multicolumn{1}{|c|}{0.64} & 
\multicolumn{1}{|c|}{2.11} & 
\multicolumn{1}{|c|}{0.60} & 
\multicolumn{1}{|c|}{8.7} & 
\multicolumn{1}{|c|}{0.60} & 
\multicolumn{1}{|c|}{0.747$\pm$0.05} \\ 
\hline
\multicolumn{1}{|c|}{0.21} & 
\multicolumn{1}{|c|}{0.66} & 
\multicolumn{1}{|c|}{7.50} & 
\multicolumn{1}{|c|}{0.80} & 
\multicolumn{1}{|c|}{9.15} & 
\multicolumn{1}{|c|}{0.69} & 
\multicolumn{1}{|c|}{0.656$\pm$0.05} \\ 
\hline
\end{tabular}
\end{center}
%\end{adjustbox}
\caption{\label{tab_S2}\textbf{Extracted values} As a measure of the normal-component resistivity for fitting the peak, we have used $\rho_{\text{n}}= 283, 200, 111, 60, 31 \mu\Omega$m, respectively. These values were obtained by fitting the data to a fourth-degree polynomial in the interval $2 T_{\text{c}} < T < 295$ K, and extrapolating down to $T_{\text{c}}$. We express $\sigma_{\text{p},0}$ in the units of anticipated BCS value $\frac{e^2}{16\hbar d}$, where $d=6.6\,$\AA{ }is the interlayer distance.}
\label{tab:measure_normalcomponent_resistivity_fitting_peak}
\end{table}

\subsubsection{Fitting of the basic model}
We can also use this simple model to find estimates for the mass anisotropies of the Cooper-pairs. By assuming a constant normal-component resistivity, we can find estimates for both $\sigma_{\text{p},0}$ and $\lambda$ by fitting the model to the height and the position of the peak in transverse response. 

Within the basic model the magnitude of the peak is given by $\rho_{\text{T},\text{peak}} =\frac{1}{2\sigma_{n,0}}(1-\lambda)/(1+\lambda)$, located at $ T_{\text{peak}} = T_{\text{c}}\left(1+\sigma_{\text{p},0}/\sigma_{\text{D},0}\right)$. We assume no normal state anisotropy, $\lambda_{\text{n}} =1$, since this mainly affects the tail of the transverse response. As an estimate of $T_{\text{c}}$, we use the temperature where the resistance drops to 1$\%$ of the value at the peak, and we take $\rho_{\text{n}} = 1/\sigma_{\text{D},0}$ to be the longitudinal response at $T_{\text{c}}$ extrapolated from a fit in the interval $2 T_{\text{c}} < T < 295$ K, see the caption of Table \ref{tab_S2}. The angles along which the samples were measured are listed in Figure \ref{fig_S6}. The results are presented in Table \ref{tab_S2} in the column labeled ‘Peak’, alongside estimates based on other approaches which we will discuss next.

\section{\label{sec_S4} Fit from low-temperature asymptotes}

Close to $T_{\text{c}}$ the conductivity is dominated by the paraconductivity. Thus, fitting to the data near $T_{\text{c}}$ should not be sensitive to the details of the normal component of conductance (which mainly affects the high-temperature tail of the data), nor to possible angular twists. According to the basic model presented in Section \ref{sec_S3.3}, both $\rho_{\text{T}}$ and $\rho_{\text{L}}$ should approach $T_{\text{c}}$ linearly in $(T-T_{\text{c}})$ with slopes that depend on $\sigma_{\text{p},0}$ and on the superconducting mass ratio $\lambda$. In fact, the quotient $\rho_{\text{T}}/\rho_{\text{L}}$ should approach a constant value that depends only on 
$\lambda$, giving a prediction for the superconducting mass anisotropy that is independent of any other parameters. This method is similar to what was presented in \cite{glover1967ideal}.

\begin{figure}[ht]
\centering
\includegraphics[width=0.9\textwidth]{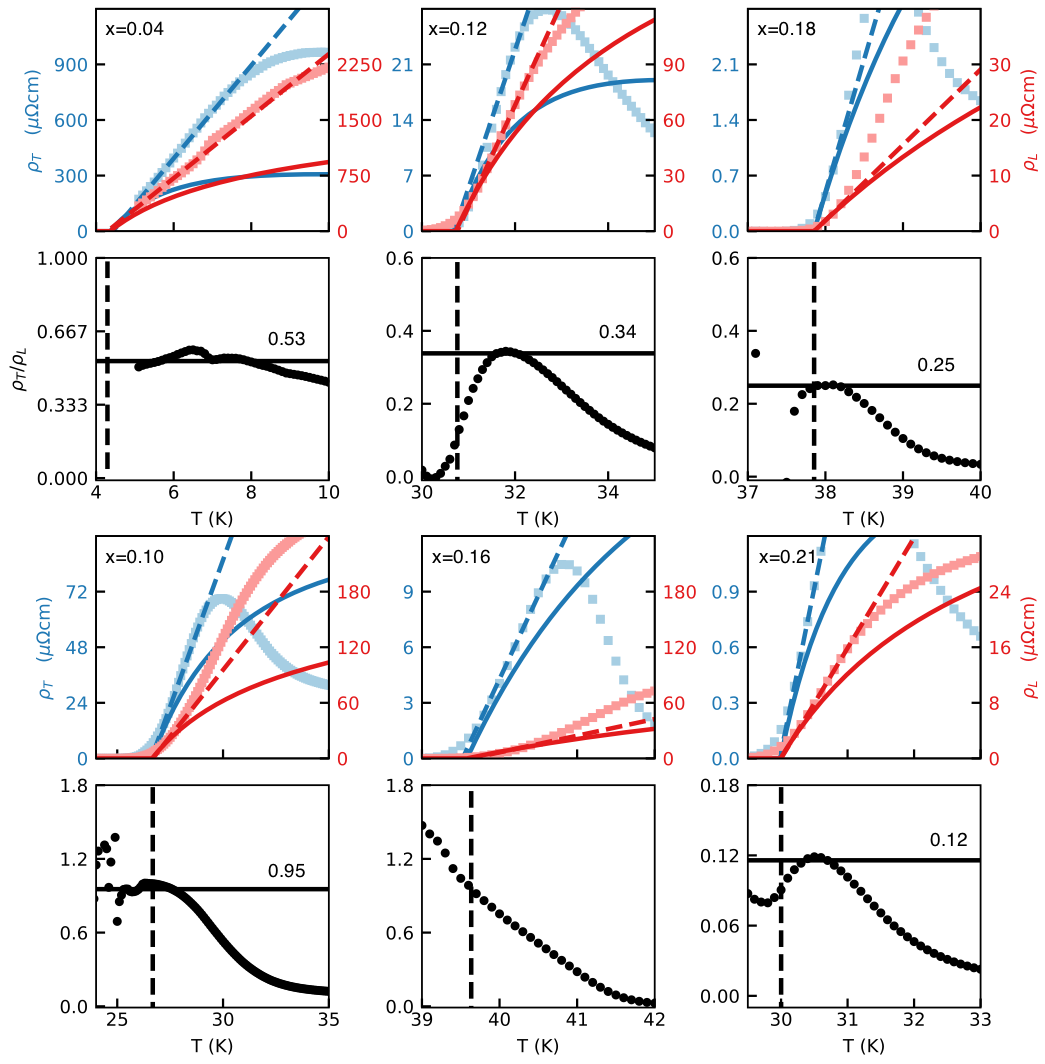}
\caption{\label{fig_S5}\textbf{Asymptotic fits.} Blue and red squares are the experimental data for transverse and longitudinal resistivity, respectively. The dashed lines correspond to the linear fit \eqref{eq:linear_fit} whereas the solid lines are the predictions based on the full model and using the extracted parameters. The black dots represent the quotient $\rho_{\text{T}}/\rho_{\text{L}}$ and the vertical dashed line is the estimated $T_{\text{c}}$ from fitting the low-temperature asymptote of $\rho_{\text{L}}$. The quotient is expected to approach $(1-\lambda^2)(1+\lambda^2)$ according to \eqref{eq:const_fit}, providing a single parameter fit of the Cooper-pair mass anisotropy $\lambda=\sqrt{m_{\text{p},y}/m_{\text{p},x}}$. (Details as discussed in text.) The extracted mass-anisotropies are shown in Table \ref{tab_S2}.}
\label{fig:fits_blue_and_red_squares}
\end{figure}

Expanding the expressions under \eqref{eq:rho_simple} near $T_{\text{c}}$ and setting $\lambda_{\text{n}} = 1$ yields

\begin{eqnarray}
\rho_{\text{T}} &  = &\frac{1/\lambda -\lambda }{2\sigma_{\text{p}}}t+\mathcal{O}(t^{2}) \nonumber\\ 
\rho_{\text{L}} &  = &\frac{1/\lambda +\lambda }{2\sigma_{\text{p}}}t+\mathcal{O}(t^{2}) 
\label{eq:linear_fit}
\end{eqnarray}

where $ t =\frac{T-T_{\text{c}}}{T_{\text{c}}}$ is the reduced temperature. In Figure \ref{fig_S5} a linear fit is shown for the available samples. The extracted parameters are shown in Table \ref{tab_S2} under \textit{Linear}. The linear fit holds reasonably well for extended parts of the curve. However, close to the anticipated 
$T_{\text{c}}$ the experimental data start to deviate from a straight line.

Expanding the quotient $\rho_{\text{T}}/\rho_{\text{L}}$ near $ T_{\text{c}}$ gives
\begin{equation}
\frac{\rho_{\text{T}}}{\rho_{\text{L}}} =\frac{1-\lambda^{2}}{1+\lambda^{2}}+\mathcal{O}\left(t\right).
\label{eq:const_fit}
\end{equation}
An estimate of this leading-order term is shown as a solid black line in Figure \ref{fig_S5}. We see that the curves level out (marked as a dashed vertical line) as $ T_{\text{c}}$ is approached, and after this they start to deviate, consistent with what was seen in the linear fit\textbf{. }In fact, this kind of deviation is expected if one assumes some spread in $ T_{\text{c}}$. If $ T_{\text{c}}$ is somewhat smeared, we can only expect the expansion to be qualitatively correct sufficiently far above the transition, but then the validity of the expansion itself becomes questionable. In Section \ref{sec_S5} we will remedy this by including a spread in $ T_{\text{c}}$ and we will refer to the resulting model as the full model. We have included the results from the full model in Figure \ref{fig_S5} for comparison. Indeed, the predictions based on the full model and the fitted parameters deviate from the experimental data in the temperature region where the linear behavior should begin. 

Nevertheless, the estimates of $\lambda$ based on this low-$T$ asymptotic fits are fairly consistent with the analysis in Section \ref{sec_S5}. The exception is for sample $ x = 0.16$ where the asymptote is not reached, and so this sample is excluded. Furthermore, as can be noted from Figure \ref{fig_S6}, the $ x = 0.10$ sample shows some discrepancy between $T_{\text{c}}$ in the longitudinal and the transverse response. This results in a leading-order term in \eqref{eq:const_fit} that is bigger than $ 1$, corresponding to an imaginary $\lambda$. However, guided by the very small $\lambda$ extracted using the methods of Section \ref{sec_S5}, together with the sensitivity to the angle $\phi$ and director $\alpha$, we suspect that these latter parameters might not have been determined accurately enough. Instead, we note that if we compensate the $ x = 0.10$ data using another angle, the leading-order term gets reasonable. In fact, the smallest possible value turns out to be $ (1-\lambda^{2})/(1+\lambda^{2})\approx 0.95$ and it occurs for $\varphi  = \phi -\alpha \approx 70^{\circ }$. This is the value used in Figure \ref{fig_S5}. This is consistent with a very high mass anisotropy of the $ p = 0.10$ sample as indicated by the more elaborate fit. However, to draw quantitative conclusions from this type of analysis, more data, over a range of angles, would be required.

These three different methods of fitting all provide consistent values for $m_{\text{p},y}/m_{\text{p},x}$ and show substantial anisotropies, especially for $x=0.10$. For the optimally-doped and overdoped samples, we find the estimates for $\sigma_{\text{p},0}$ that are substantially higher than $\frac{e^{2}}{16\hbar d}$, the anticipated BCS value (see Section \ref{sec_S1}). As discussed in Section \ref{sec_S1}, one should in general expect deviations from the value $\frac{e^{2}}{16\hbar d}$, since the relaxation dynamics of the Cooper-pairs in cuprates may indeed be different than what is expected in the BCS theory. One other reason for overestimation of $\sigma_{\text{p},0}$is that in the overdoped regime, where the conductivity is quite high, one needs to take into account the temperature dependence of the normal conductivity. Next we will discuss the already mentioned full model that takes into account a non-trivial temperature dependence of the normal conductivity and a normal electron mass anisotropy. 

\section{ \label{sec_S5} Extended model}

The fits presented in Section \ref{sec_S3} and \ref{sec_S4} utilize the properties of the conductivity just above $T_{\text{c}}$ and can therefore provide a simple measure of the paraconductivity strength $\sigma_{\text{p},0}$ and the anisotropy of pair mass. The advantage of this approach is that the fits are relatively insensitive to the details of the normal electron response and to the twist of the principal axis of the conductivity. 

Here, instead of fitting to the response exclusively near $T_{\text{c}}$, we match the longitudinal and transverse responses at all temperatures. By having a model that predicts the complete temperature dependence we can address the quality of the fit, assessing the goodness of our model (see Section \ref{sec_S5.4}). Furthermore, we will relax the assumption that the normal state is isotropic. This gives us the possibility of explaining the observed twist in $\alpha $ (for the $x=0.16$ sample) by assuming that the principal axes of $\overline{\sigma }_{\text{n}}$ and $\overline{\sigma }_{\text{p}}$ are not aligned. 

First, we extend the expression \eqref{eq:sig_simple} to include the case when superconducting and normal component are not aligned with one another. Expressing the conductivity in the superconducting principal frame (aligned with $\overline{\sigma }$ at $T_{\text{c}}$) oriented with an angle relative to the crystallographic $a$-axis (the [100] direction) gives

\begin{equation}
\overline{\sigma }_{\text{p}} =\begin{bmatrix}
\sigma_{\text{p},x} & 0 \\ 
0 & \sigma_{\text{p},y} \\ 
\end{bmatrix},
\end{equation}

\begin{equation}
\overline{\sigma }_{\text{n}} = \sigma_{\text{D}}\begin{bmatrix}
\cos(\nu)  & -\sin(\nu) \\ 
\sin(\nu) & \cos(\nu) \\ 
\end{bmatrix}\begin{bmatrix}
\lambda_{\text{n}} & 0 \\ 
0 & 1/\lambda_{\text{n}} \\ 
\end{bmatrix}\begin{bmatrix}
\cos(\nu)  & \sin(\nu) \\ 
-\sin(\nu) & \cos(\nu) \\
\end{bmatrix}
\end{equation}

where $\nu=\alpha\left(295K\right)-\alpha(T_{\text{c}})$ is the relative angle of the normal component principal frame (see Table \ref{tab_S1}). (We will extend this further to include non-zero magnetic field in Section \ref{sec_S5.2}.

To make as few assumptions as possible, we fit the temperature dependence of the normal resistivity to the high-temperature asymptote of the longitudinal response,  $\rho_{\text{L}}\left(T\right)\rightarrow \rho_{\text{L},\text{Asym.}}(T)$, given by

\begin{equation}
\rho_{\text{L},\text{Asym.}}\left(T\right) =\frac{\frac{1}{\lambda_{\text{n}}}+\lambda_{\text{n}}}{2\sigma_{\text{D}}\left(T\right)}.
\label{eq:asym_rho}
\end{equation}

Here we have only taken into account the rotationally-invariant part that dominates the longitudinal response. Next, we fit $\rho_{\text{L},\text{Asym.}}(T)$ to a third-degree polynomial $(a+bT+cT^2+dT^3)$ for $x=0.10$, 0.12, 0.16, 0.18 and 0.21, respectively, over an interval ranging from $2T_{\text{c}}$ (30 K for $x=0.04$) to $290\,$K, with a resolution of $0.1\,$K. For a given $\lambda_{\text{n}}$, we can solve \eqref{eq:asym_rho} for
$\sigma_{\text{D}}(T)$, attributing the temperature dependence to the relaxation time $\tau$, in $\sigma_{\text{D}} =\frac{ne^{2}\tau }{\sqrt{m_{\text{n},y}m_{\text{n},x}}}$. In the $x=0.04$ sample, the longitudinal resistivity increases at low temperature, indicating an incipient transition into an insulating state, so we include a term $1/T$ in the fit, i.e., we fit $\rho_{\text{L},\text{Asym.}}(T)$ to $(a/T+b+cT+dT^2$ over the interval 30 K to 290 K. Since the measured $\rho_{\text{L}}$ inevitably includes paraconductivity effects even at temperatures above $2T_{\text{c}}$ (or 30 K for $x=$ 0.04), adding the paraconductivity contribution on top would exaggerate the para-conducting properties. To compensate for this, we include a rescaling parameter, $r$, of the most singular term $a \rightarrow ar$, to the fit (with $ r = 1$ as the initial guess).

As a final step, we model the sample as consisting of serially-coupled domains from a Gaussian distribution with the mean temperature $ T_{\text{c}}^{\text{m}}$ and the standard deviation $\Delta T_{\text{c}}$, to account for the width of the peak in the transverse resistivity data

\begin{equation}
\rho_{\text{T},\text{model}}\left(\phi,T\right) =\frac{1}{\Delta T_{\text{c}}\sqrt{2\pi }}\int_{-\infty }^{\infty }dT_{\text{c}}\,e^{-\frac{(T_{\text{c}}-T_{\text{c}}^{\text{m}})^{2}}{2\Delta T_{\text{c}}^{2}}}\rho_{\text{T}}\left(\phi,T_{\text{c}},T\right),
\label{eq:full_fit}
\end{equation}

where $\rho_{\text{T}}(\phi,T_{\text{c}},T)$ is the transverse resistivity for a single $ T_{\text{c}}$. An analogous expression was used for the longitudinal response. The transition width modeled in this way is found to be consistent with the mutual inductance measurements (MI) that provide a measure of the spread in $ T_{\text{c}}$ (see Table \ref{tab_S3}). 

\subsection{The results --- zero field}

\begin{figure}[ht]
\centering
\includegraphics[width=0.9\textwidth]{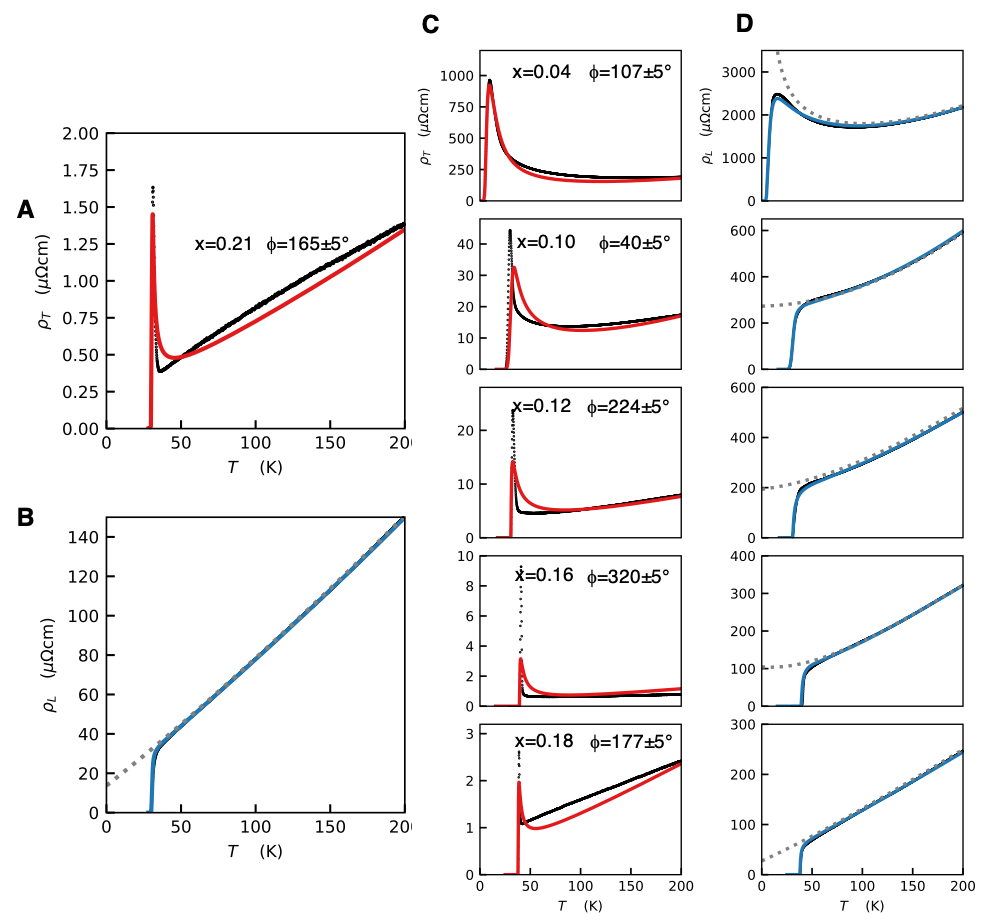}
\caption{\label{fig_S6}\textbf{Modelling temperature and doping dependence of longitudinal and transverse resistivity of LSCO.} A, Red line: predicted $\rho_{\text{T}}(T)$; black line: experimental data. B, Blue line: predicted $\rho_{\text{L}}(T)$ dependence; black line: experimental data. The model utilizes a normal component response $\rho_{\text{n}}(T)$ (the dashed gray line) obtained by extrapolating from the high-temperature data fit. For doping $x=0.21$. C, The same as in A, for $x=$ 0.04, 0.10, 0.12, 0.16, and 0.18, respectively. D, The same as in B, for $x=$ 0.04, 0.10, 0.12, 0.16, and 0.18.} 
\label{fig:temperature_and_doping_dependence_longitudinal}
\end{figure}

The results of the fits to the full temperature dependences are shown in Figure \ref{fig_S6}. The model captures well the temperature dependence of both the longitudinal and the transverse resistivity, notably reproducing the pronounced peak in $\rho_{\text{T}}(T)$ near $T_{\text{c}}$. For some doping levels the model seemingly produces worse fit than for others. However, it should be noted that the discrepancy is within the expected value given the measured noise (see Section \ref{sec_S5.4}). 

The mass ratios extracted by fitting the data are shown in Figure \ref{fig_2}A in the main text and in Table \ref{tab_S2} (along the other estimates) and Table \ref{tab_S3}). The mass anisotropy forms a parabola, similar to that of the $T_{\text{c}}(x)$ dependence. The anisotropy of the superconducting fluctuations sharply increases on the underdoped side, while the anisotropy of the normal-state quasiparticles remains relatively weak for all doping levels.

\subsection{\label{sec_S5.2} Finite magnetic field --- results}

In the presence of magnetic field, we only consider the $x=$ 0.16 data, since the twist with temperature is most pronounced at this doping. Again, to account for the rotation of the principal axes of conductivity, we align the normal component mass tensor with the high-temperature director and the superconducting mass tensor with the low-temperature director. In Figure \ref{fig_3} in the main text, we show the simulated results for 
$\rho_{\text{L}}\left(T\right)$ and $\rho_{\text{T}}(T)$, for a range of $ B$ using the parameter values from the fit of the $B=0$ data for the $x=0.16$ sample: $\Delta T_{\text{c}}\left(0\right)=0.29\,$K, $\beta =0.34$ (see \eqref{eq:broadening}), and $\frac{m_{\text{p},y}}{m_{\text{p},x}}  =  0.765$. 

To model the temperature dependence of the normal conductivity, we use a linear temperature fit to the $B=$ 0 longitudinal resistivity (in Figure \ref{fig_1}D), over the interval $45\,$K$< T < 72\,$ K, with the normal mass ratio $m_{\text{n},y}/m_{\text{n},x}=0.97$. The longitudinal response (Figure \ref{fig_3}A) accounts well for the measured data (Figure \ref{fig_1}D), while the transverse response yields qualitatively correct behavior. For small $B$, we reproduce the expected sharp twist of the nematic director near $T_{\text{c}}$, compared to its orientation at high temperatures. As the field is increased, the transition is broadened and pushed to lower temperatures, effectively yielding a gradual rotation of the nematic director as a function of B at a fixed temperature and a suppression of the peak. This is consistent with the fixed-T and -B experimental snapshots of this behavior in Figure \ref{fig_1}B,C in the main text.

\subsection{Detailed fitting procedure}

The modeled prediction in Figure \ref{fig_S6} was fitted to the transverse and longitudinal data, by tuning  $ T_{\text{c}}^{\text{m}},\Delta T_{\text{c}},\lambda,\lambda_{\text{n}}$ and $r$ to minimize the root-mean-square error using the Lmfit-py package\footnote{https://lmfit.github.io/lmfit-py/index.html}. Since the longitudinal response is insensitive to $\lambda,\lambda_{\text{n}}$, while the transverse response is insensitive to $r$, the fit was split into two parts. First, $r$ was fitted to the longitudinal resistivity over the interval $ 0.5T_{\text{c}}<T<200\,$K with a step size of $0.1\,$K, while keeping $\lambda,\lambda_{\text{n}},T_{\text{c}}^{\text{m}}$ and $\Delta T_{\text{c}}$ constant. Then, $ T_{\text{c}}^{\text{m}}, \Delta T_{\text{c}}$, and the anisotropy-measures $\lambda$ and $\lambda_{\text{n}}$ were fitted to the transverse response, over the interval  $ 0.5 T_{\text{c}}<T<1.3 T_{\text{c}}$ (30 K for $x=$ 0.04), with a step size of 0.1 K, and over the interval $ 1.3T_{\text{c}}<T<290\,$K with a step size of 1.0 K, while keeping $r$ fixed. This procedure was iterated over (using the previously fitted parameters as an improved guess) until the parameter values converged to within the uncertainty of one standard deviation of the fit itself. The resulting fits are plotted in Figure \ref{fig_S6} and the parameter values are listed in Table \ref{tab_S3}). In simulating the data, the normal and superconducting directors are assumed to be aligned for all samples, except for $x=$ 0.16, where the relative angle of 60 degrees between the two was used (see Table \ref{tab_S1}).

In Section \ref{sec_S3} and \ref{sec_S4} we discussed fits which included $\sigma_{\text{p},0}$ as well as the pair-mass anisotropy, presented in Table \ref{tab_S2}. We noted that these are generally bigger than the BCS prediction but mentioned that this estimate should decrease when the normal conductivity is better accounted for. Indeed, by including a normal conductivity in the way described in \eqref{eq:asym_rho} and fitting to 
$\sigma_{\text{p},0}$ in addition to $ T_{\text{c}}^{\text{m}},  \Delta T_{\text{c}},  \lambda,  \lambda_{\text{n}}$, and $r$ yields a lower estimate of $\sigma_{\text{p},0}$, on the order of $0.5$--$1.5$ (in the units of
$\frac{e^{2}}{16\hbar d}$ where $d=$ 6.6{\AA} denotes the La\textsubscript{2-x}Sr\textsubscript{x}CuO\textsubscript{4} inter-layer distance). However, to obtain the parameters in Table \ref{tab_S3} the parameter $\sigma_{\text{p},0}$ was fixed to its BCS predicted value. The reason for leaving out $\sigma_{\text{p},0}$ of the fit is because there is a very high correlation between $\sigma_{\text{p},0}$ and $ T_{\text{c}}^{\text{m}},  \Delta T_{\text{c}},  \lambda$, when fitting to the transverse response, as well as between $\sigma_{\text{p},0}$ and $r$ when fitting to the longitudinal response. It is therefore undesirable to include a fit to $\sigma_{\text{p},0}$ when $\lambda$ is our primary parameter of interest. Nevertheless, the overall conclusion that the mass-anisotropy  $\lambda$  must be high turns out to be insensitive to $\sigma_{\text{p},0}$.

Table \ref{tab_S3} also includes an estimate for the spread in $ T_{\text{c}}$ from mutual inductance (MI) measurements on the films before lithography patterning. These values are with the Gaussian distribution of critical temperatures presented in \eqref{eq:full_fit}. The MI measurements probe a much larger area ($5$--$10\,$ mm) of the film than that of a device ($300\,\mu$m) used in transport measurements, so we expect that MI should somewhat overestimate $\Delta T_{\text{c}}$. The methods of fit discussed in Section \ref{sec_S3} and \ref{sec_S4} all assume $\Delta T_{\text{c}} = 0$. Thus, the MI measurements are in favor of the more elaborate model presented here.

\subsection{\label{sec_S5.4} Estimated error}

The one-standard-deviation error of the parameters estimated from fitting the transverse and longitudinal response for one Hall-bar device is very small. However, this neglects the systematic errors in the fit coming from the uncertainty of the angle and also the device-to-device variations due to lithography. The errors originating from the $\pm5^\circ$ angle uncertainty was estimated by varying the observation angle in the fit. To properly account for the device-to-device variations, we would need an ensemble of bars at different temperatures, which is not available to us. Instead, as a measure of how the device variance propagates to an uncertainty in the estimates of $ m_{\text{p},y}/m_{\text{p},x}$ and $ m_{\text{n},y}/m_{\text{n},x}$, we fitted a rescaled transverse resistivity $\rho'_{T}(\phi,T)= R\rho_{\text{T}}(\phi,T)$ where $R$ is a number given by

\begin{equation}
R \rho_{\text{T}}\left(\phi,T = 295K\right) =  \rho_{\text{T}}\left(\phi,T = 295K\right)+ \sigma_{\text{bar}}(295K)
\label{eq:rescale}
\end{equation}
where $\sigma_{\text{bar}}$(295 K) is the standard deviation inferred from the angular measurement, see Figure \ref{fig_S3}. These three errors are listed in Table \ref{tab_S4} and the biggest constitute the error bars in Figure \ref{fig_S6}.

\begin{table}[ht]
\footnotesize
\flushleft
\begin{adjustbox}{max width=\textwidth}
\begin{tabular}{p{0.7cm}p{4.4cm}p{4.4cm}p{3.93cm}p{3.57cm}p{1.5cm}}
\hline
\multicolumn{1}{|c|}{x} & 
\multicolumn{1}{|c|}{$ m_{\text{p},y}/m_{\text{p},x}$} & 
\multicolumn{1}{|c|}{$ m_{\text{n},y}/m_{\text{n},x}$} & 
\multicolumn{1}{|c|}{$ T_{\text{c}}$(K) (Fitted)} & 
\multicolumn{1}{|c|}{$\Delta T_{\text{c}}$ (K) (Fitted)} & 
\multicolumn{1}{|c|}{$ T_{\text{c}}$ (MI)} \\ 
\hline
\multicolumn{1}{|c|}{{\footnotesize 0.04}} & 
\multicolumn{1}{|c|}{{\footnotesize 0.2070$\pm$0.001(0.02)[0.04]}} & 
\multicolumn{1}{|c|}{{\footnotesize 0.8650$\pm$0.001(0.003)[0.02]}} & 
\multicolumn{1}{|c|}{{\footnotesize 5.360$\pm$0.02(0.07)[0.095]}} & 
\multicolumn{1}{|c|}{{\footnotesize 0.810$\pm$0.06(0.09)}} & 
\multicolumn{1}{|c|}{-} \\ 
\hline
\multicolumn{1}{|c|}{{\footnotesize 0.10}} & 
\multicolumn{1}{|c|}{{\footnotesize 0.128$\pm$0.001(0.08)[0.11]}} & 
\multicolumn{1}{|c|}{{\footnotesize 0.941$\pm$0.001(0.009)[0.003]}} & 
\multicolumn{1}{|c|}{{\footnotesize 27.3$\pm$0.04(0.015) [0.01]}} & 
\multicolumn{1}{|c|}{{\footnotesize 0.47$\pm$0.003(0.36)}}  & 
\multicolumn{1}{|c|}{{\footnotesize 0.97}} \\ 
%\hline
\multicolumn{1}{|c|}{} & 
\multicolumn{1}{|c|}{} & 
\multicolumn{1}{|c|}{} & 
\multicolumn{1}{|c|}{{\footnotesize 30.0$\pm$0.02(0.318) [0.65]}}  & 
\multicolumn{1}{|c|}{{\footnotesize 1.66$\pm$0.001(0.54)}}  & 
\multicolumn{1}{|c|}{} \\ 
\hline
\multicolumn{1}{|c|}{{\footnotesize 0.12}} & 
\multicolumn{1}{|c|}{{\footnotesize 0.553$\pm$0.006(0.03)[0.06]}} & 
\multicolumn{1}{|c|}{{\footnotesize 0.977$\pm$0.0006(0.0025)[0.004]}} & 
\multicolumn{1}{|c|}{{\footnotesize 30.8$\pm$0.09(0.004) [0.001]}} & 
\multicolumn{1}{|c|}{{\footnotesize 0.08$\pm$0.02(0.02)}} & 
\multicolumn{1}{|c|}{{\footnotesize 0.90}} \\ 
\hline
\multicolumn{1}{|c|}{{\footnotesize 0.16}} & 
\multicolumn{1}{|c|}{{\footnotesize 0.765$\pm$0.007(0.07)[0.20]}} & 
\multicolumn{1}{|c|}{{\footnotesize 0.965$\pm$0.00038(0.001)}} & 
\multicolumn{1}{|c|}{{\footnotesize 39.4$\pm$0.077(0.084) [0.21]}} & 
\multicolumn{1}{|c|}{{\footnotesize 0.03$\pm$0.23(0.076)}} & 
\multicolumn{1}{|c|}{{\footnotesize 0.25}} \\ 
\hline
\multicolumn{1}{|c|}{{\footnotesize 0.18}} & 
\multicolumn{1}{|c|}{{\footnotesize 0.747$\pm$0.004(0.03)[0.05]}} & 
\multicolumn{1}{|c|}{{\footnotesize 0.979$\pm$0.0001(0.003)[0.005]}} & 
\multicolumn{1}{|c|}{{\footnotesize 38.0$\pm$0.03(0.004) [0.004]}} & 
\multicolumn{1}{|c|}{{\footnotesize 0.13$\pm$0.005(0.08)}} & 
\multicolumn{1}{|c|}{{\footnotesize 0.90}} \\ 
\hline
\multicolumn{1}{|c|}{{\footnotesize 0.21}} & 
\multicolumn{1}{|c|}{{\footnotesize 0.656$\pm$0.005(0.002)[0.05]}} & 
\multicolumn{1}{|c|}{{\footnotesize 0.985$\pm$0.0001(0.0002)[0.003]}} & 
\multicolumn{1}{|c|}{{\footnotesize 30.4$\pm$0.020(0.0014) [0.007]}} & 
\multicolumn{1}{|c|}{{\footnotesize 0.29$\pm$0.03(0.006)}} & 
\multicolumn{1}{|c|}{{\footnotesize 0.33}} \\ 
\hline
\end{tabular}
\end{adjustbox}
\caption{\label{tab_S3}\textbf{Parameters from modelling the doping dependence of longitudinal and transverse resistivity of LSCO.} Three different errors are presented. The first is one standard deviation inferred from one sample. The second error, in parenthesis, refers to a shift in $\phi$ by $\pm$5 degrees. The third error, in brackets, is obtained from fitting a rescaled transverse resistivity (see \eqref{eq:rescale}). The fitted width of the transition temperature $\Delta T_{\text{c}}$ is presented alongside the spread inferred from the mutual inductance measurements (MI). For the $x=$ 0.10 sample, two independent values of $ T_{\text{c}}$ and $\Delta T_{\text{c}}$ were used for the transverse and longitudinal response, respectively. The error bar on $ T_{\text{c}}$ presented in Figure \ref{fig_2}A for $x=$ 0.10 refers to this difference. For $x=$ 0.16 another dataset was used to determine $ m_{\text{n},y}/m_{\text{n},x}$ (presented in Figure \ref{fig_S7}).
The error bars for $m_{\text{p},y}/m_{\text{p},x}$ and $ m_{\text{n},y}/m_{\text{n},x}$ presented in Figure \ref{fig_S6} refer to the biggest error obtained.}
\end{table}

\begin{figure}[ht]
\centering
\includegraphics[width=0.9\textwidth]{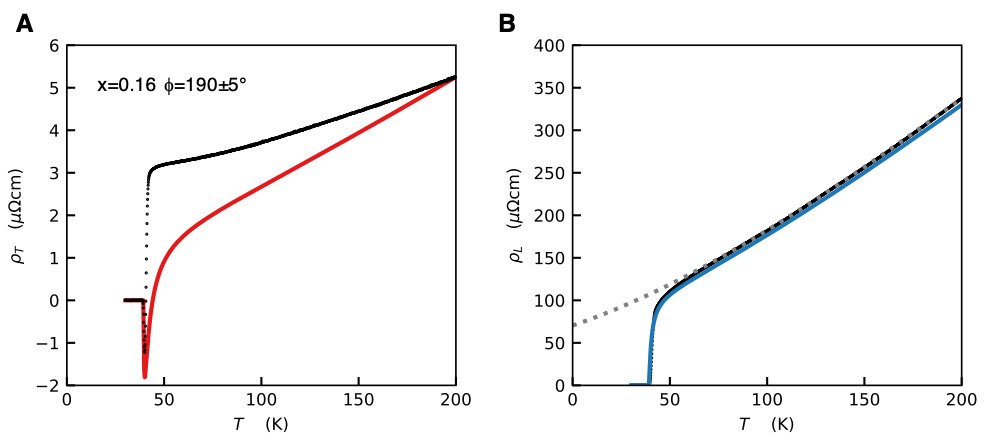}
\caption{\label{fig_S7}\textbf{Additional $\mathbf{x=0.16}$ sample to determine  $\mathbf{m}_{\mathbf{ny}}\mathbf{/}\mathbf{m}_{\mathbf{nx}}$ in Table \ref{tab_S3}}. The data for the Hall-bar
device at $\phi=$ 320$^\circ$ $\pm$ 5$^\circ$, presented in Figure \ref{fig_S6}, are unsuitable to determine $m_{y}^{\ast }/m_{x}^{\ast }$ since the director is aligned (within the measurement accuracy) with the nodal direction of the transverse resistivity at high temperatures (located at 322$^\circ$). Instead, the transverse response of the device oriented along $\phi=$ 190$^\circ$ $\pm$ 5$^\circ$, shown in A, was used to determine $ m_{\text{n},y}/m_{\text{n},x}$. In order for the original $\phi=$ 320$^\circ$ $\pm$ 5$^\circ$ device to correspond to the obtained ratio $ m_{\text{n},y}/m_{\text{n},x}=$ 0.965, the angle $\phi=$ 324$^\circ$ was used in that simulations. The measured transverse and longitudinal response for the $\phi=$ 190$^\circ$ $\pm$ 5$^\circ$ device are shown alongside the simulated response in A and B. Note that the temperature dependence in $\alpha$ here induces a change in sign of the transverse response. Fitting this curve is therefore very sensitive to the exact functional form of the twist. Still, the same parameters were used here as the response in Figure \ref{fig_S6} the $\phi=$ 320$^\circ$ $\pm$ 5$^\circ$ device.}
\end{figure}

\subsection{Goodness of fit}

To assess the quality of fit we calculated the chi-squared values, presented in Table \ref{tab_S4}, using the standard deviation $\sigma_{\text{bar}}$ as the measured error. The probability of obtaining a worse fit (see Table \ref{tab_S4}) was estimated by comparing with a chi-squared distribution of one degree of freedom. (Using the one degree of freedom distribution assumes that the errors at different temperatures, but for the same sample, are perfectly correlated, which is an oversimplification. Regardless, this yields a lower bound of $P_n(\chi^2>\chi^2_{\text{red}})$, and could therefore be seen as the most conservative estimate.) From this we conclude that our model is a good candidate for explaining the observed data.

\begin{table}[ht]
\footnotesize
\centering
%\begin{adjustbox}{min width=\textwidth}
\begin{tabular}{p{2.44cm}p{4.29cm}p{5.14cm}p{1.89cm}p{3.0cm}}
\hline
\multicolumn{1}{|c|}{$x$} & 
\multicolumn{1}{|c|}{ $\sigma_{\text{bar}}(Tc)$ ($\mu\Omega$cm)}  & 
\multicolumn{1}{|c|}{ $\sigma_{\text{bar}}(295\,$K) ($\mu\Omega$cm)}  & 
\multicolumn{1}{|c|}{ $\chi^2_{\text{red}}$ (Fit)} & 
\multicolumn{1}{|c|}{$P_1(\chi^2>\chi^2_{\text{red}})$} \\ 
\hline
\multicolumn{1}{|c|}{0.04} & 
\multicolumn{1}{|c|}{3150} & 
\multicolumn{1}{|c|}{745} & 
\multicolumn{1}{|c|}{0.35} & 
\multicolumn{1}{|c|}{55$\%$} \\ 
\hline
\multicolumn{1}{|c|}{ 0.10} & 
\multicolumn{1}{|c|}{65.9} & 
\multicolumn{1}{|c|}{16.3\ \ \ \ } & 
\multicolumn{1}{|c|}{0.15} & 
\multicolumn{1}{|c|}{70$\%$} \\ 
\hline
\multicolumn{1}{|c|}{ 0.12} & 
\multicolumn{1}{|c|}{14.7} & 
\multicolumn{1}{|c|}{4.3} & 
\multicolumn{1}{|c|}{0.40} & 
\multicolumn{1}{|c|}{53$\%$} \\ 
\hline
\multicolumn{1}{|c|}{ 0.16} & 
\multicolumn{1}{|c|}{1.51} & 
\multicolumn{1}{|c|}{0.60} & 
\multicolumn{1}{|c|}{0.63} & 
\multicolumn{1}{|c|}{43$\%$} \\ 
\hline
\multicolumn{1}{|c|}{ 0.18} & 
\multicolumn{1}{|c|}{0.79} & 
\multicolumn{1}{|c|}{0.50} & 
\multicolumn{1}{|c|}{0.048} & 
\multicolumn{1}{|c|}{83$\%$} \\ 
\hline
\multicolumn{1}{|c|}{ 0.21} & 
\multicolumn{1}{|c|}{0.22} & 
\multicolumn{1}{|c|}{0.29} & 
\multicolumn{1}{|c|}{0.047} & 
\multicolumn{1}{|c|}{83$\%$} \\ 
\hline
\end{tabular}
%\end{adjustbox}
\caption{\label{tab_S4}\textbf{Error of fits.} Sample spread $\sigma_{\text{bar}}$ in $\rho_{\text{T}}$ estimated from the angular fit shown in Figure \ref{fig_S3} at $T_{\text{c}}$ and at 295 K. These were used as the measured errors when assessing the fitted model $\rho_{\text{T},\text{model}}\left(\phi,T\right)$. The resulting reduced chi-square value (employing a linear interpolation between the low- and high-temperature value) of the corresponding fits are listed together with the probability of obtaining a worse fit, from a chi-squared distribution of one freedom. }
\end{table}

\end{document}